\documentclass[12pt]{article}

\ifx\pdfoutput\undefined
\usepackage[dvips,bookmarks]{hyperref}
\else
\usepackage{hyperref}
\fi
\hypersetup{colorlinks=false,bookmarksopen,bookmarksnumbered,citecolor=blue,
   pdfstartview=FitH}

\usepackage[dvips]{graphicx}
\usepackage{latexsym}
\usepackage{amssymb,amsfonts,amsmath}
\usepackage{graphicx} 
\usepackage{indentfirst}

 \usepackage{bbm}

\parskip=\medskipamount

\newcommand {\cD}{{\cal D}}
\newcommand {\cE}{{\cal E}}

\newcommand {\cL}{{\cal L}}
\newcommand {\cM}{{\cal M}}
\newcommand {\cN}{{\cal N}}
\newcommand {\cO}{{\cal O}}
\newcommand {\cP}{{\cal P}}

\newcommand {\cR}{{\cal R}}

\newcommand {\cT}{{\cal T}}

\newcommand {\cV}{{\cal V}}

\def\a{\alpha}
\def \bi{\bibitem}

\def\b{\beta}

\def\d{\delta}

\def\g{\gamma}

\def\l{\lambda}
\def\m{\mu}
\def\n{\nu}
\def\o{\omega}
\def\p{\pi}
\def\q{\theta}

\newcommand{\qb}{{\bar{\theta}}}

\def\r{\rho}
\def\s{\sigma}

\def\z{\zeta}
\def\D{\Delta}
\def\F{\Phi}

\def\O{\Omega}

\def\S{\Sigma}

\def\rd{{\rm d}}
\def\ri{{\rm i}}

\newcommand{\ad}{{\dot{\alpha}}}                           %new
\newcommand{\bd}{{\dot{\beta}}}                            %new
\newcommand{\ve}{\varepsilon}                            %new
\newcommand{\cDB}{{\bar\cD}}                            %new
\newcommand{\DB}{\bar{D}}

\newcommand{\pa}{\partial}                           %new
\newcommand{\hf}{\frac12}
\newcommand{\be}{\begin{equation}}
\newcommand{\ee}{\end{equation}}
\newcommand{\bea}{\begin{eqnarray}}
\newcommand{\eea}{\end{eqnarray}}
\newcommand{\non}{\nonumber}
\newcommand{\1}{{\underline{1}}}
\newcommand{\2}{{\underline{2}}}

\def\dt#1{{\buildrel {\hbox{\LARGE .}} \over {#1}}}    % dot-over for sp/sb

\newcommand{\bm}[1]{\mbox{\boldmath$#1$}}

\def\double #1{#1{\hbox{\kern-2pt $#1$}}}

\newcommand{\hm}{{\hat{m}}}

\newcommand{\ha}{{\hat{a}}}
\newcommand{\hb}{{\hat{b}}}

\newcommand{\hal}{{\hat{\a}}}
\newcommand{\hbe}{{\hat{\b}}}
\newcommand{\hga}{{\hat{\g}}}
\newcommand{\hde}{{\hat{\d}}}
\newcommand{\hrh}{{\hat{\rho}}}
\newcommand{\hta}{{\hat{\tau}}}

\newcommand{\gd}{{\dot\g}}
\newcommand{\dd}{{\dot\d}}

\newcommand{\ts}{{\tilde{\s}}}

\newcommand{\CD}{{\nabla}}

\newcommand{\teb}{{\bar{\theta}}}

\newcommand{\upm }{{(u^+u^-)}}

\begin{document}
%%%%%%%%%%%%%%%%
%%%%%%%%%%%%%%%%

\begin{titlepage}

\begin{flushright}
UMDEPP 08-027\\
December, 2008\\
\end{flushright}
\vspace{5mm}

\begin{center}
{\Large \bf Different representations for the action principle }\\
{\Large \bf in 4D $\bm{\cN=2}$  supergravity}
\end{center}

\begin{center}

{\large  
Sergei M. Kuzenko\footnote{{kuzenko@cyllene.uwa.edu.au}}${}^{a}$
and 
Gabriele Tartaglino-Mazzucchelli\footnote{gtm@umd.edu}${}^{a,b}$
}\\
\vspace{5mm}

\footnotesize{
${}^{a}${\it School of Physics M013, The University of Western Australia\\
35 Stirling Highway, Crawley W.A. 6009, Australia}}  
~\\
\vspace{2mm}

${}^{b}${\it Center for String and Particle Theory,
Department of Physics \\
University of Maryland, College Park, MD 20742-4111, USA}
~\\

\end{center}
\vspace{5mm}

\begin{abstract}
\baselineskip=14pt
Within the superspace formulation for
four-dimensional $\cN=2$ matter-coupled supergravity 
developed in arXiv:0805.4683, we elaborate two approaches to reduce the superfield 
action to components. One of them is based on the principle of projective 
invariance which is a purely $\cN=2$ concept having no analogue in simple supergravity.
In this approach, the component reduction of the action is performed {\it without} imposing 
any Wess-Zumino gauge condition, that is by keeping intact  all the gauge symmetries
of the superfield action, including the super-Weyl invariance. 
As a simple application, the c-map is derived for the first time from  superfield supergravity.
Our second approach to component reduction is based on the method of normal 
coordinates around a submanifold in a curved superspace, which we develop in detail.
We derive differential equations which are obeyed by the vielbein and the connection 
in normal coordinates,
and which can be used to reconstruct these objects, in principle in closed form. 
A separate equation is found for the super-determinant of the vielbein $E= {\rm Ber} (E_M{}^A)$, 
which allows one to reconstruct $E$ without a detailed knowledge of the vielbein.
This approach is applicable to any supergravity theory in any number of space-time dimensions.
As a simple application of this construction, we reduce an integral over the 
curved $\cN=2$ superspace to that over the chiral subspace of the full superspace. 
We also give a new representation for the curved projective-superspace action principle 
as a chiral integral.
\end{abstract}
\vspace{1cm}

\vfill
\end{titlepage}

\newpage
\renewcommand{\thefootnote}{\arabic{footnote}}
\setcounter{footnote}{0}

\tableofcontents{}
\vspace{1cm}
\bigskip\hrule

%%%%%%%%%%%%%%%%%%%%%%%%%%%%%%%%%%%%%%%%%%%%%%%%%%
%%%%%%%%%%%%%%%%%%%%%%%%%%%%%%%%%%%%%%%%%%%%%%%%%%

\section{Introduction}
\setcounter{equation}{0}
One of the main virtues of superspace approaches to supergravity theories 
in diverse dimensions is the possibility 
to write down the most general  locally supersymmetric actions formulated in terms 
of a few superfield dynamical variables possessing, as a rule,  a transparent geometric origin. 
The price to pay for this generality is that working out a reduction from the parental superfield action 
to its component counterpart requires some special care. Being trivial conceptually, such a 
reduction may be technically quite involved and challenging.

The present paper is aimed at carrying out a component reduction, 
as well as a partial superspace reduction, for the action principle occurring within 
the superspace formulation for four-dimensional $\cN=2$ matter-coupled supergravity 
recently developed in \cite{KLRT-M},  as a natural extension of  
the earlier construction for 5D $\cN=1$  supergravity \cite{KT-Msugra1,KT-Msugra3}.
The  matter fields in \cite{KLRT-M} are described in terms of covariant projective 
multiplets which are curved-space versions of  the superconformal 
projective multiplets \cite{K-hyper2} living in rigid projective superspace
\cite{LR}. In addition to the local $\cN=2$ superspace
coordinates\footnote{World indices take values
$m=0,1,\cdots,3$, $\mu=1,2$, $\dot{\mu}=1,2$ and  $i=\1,\2$,
and similarly for tangent space indices;
see Appendix \ref{SCG} for our notation and conventions.}
$z^{{M}}=(x^{m},\q^{\mu}_i,{\bar \q}_{\dot{\mu}}^i)$,
such a supermultiplet, $Q^{(n)}(z,u^+)$, depends on auxiliary  
isotwistor  variables $u^{+}_i \in  {\mathbb C}^2 \setminus  \{0\}$, 
with respect to which $Q^{(n)}$ is holomorphic and homogeneous,
$Q^{(n)}(c \,u^+) =c^n \,Q^{(n)}(u^+)$, on an open domain of ${\mathbb C}^2 \setminus  \{0\}$
(the integer parameter $n$ is called the weight of   $Q^{(n)}$).
In other words, such superfields are intrinsically defined in  ${\mathbb C}P^1$.
The covariant projective  supermultiplets are required to be 
annihilated by half of the supercharges, 
\be
\cD^+_{\a} Q^{(n)}  = {\bar \cD}^+_{\ad} Q^{(n)}  =0~, \qquad \quad 
\cD^+_{ \a}:=u^+_i\,\cD^i_{ \a} ~, \qquad
{\bar \cD}^+_{\dot  \a}:=u^+_i\,{\bar \cD}^i_{\dot \a} ~,
\label{ana-introduction}
\ee  
with $\cD_{{A}} =(\cD_{{a}}, \cD_{{\a}}^i,\cDB^\ad_i)$ the covariant superspace 
derivatives. The dynamics of supergravity-matter 
systems are described by locally supersymmetric actions of the form \cite{KLRT-M}:
\bea
S&=&
\frac{1}{2\pi} \oint_C (u^+ \rd u^{+})
\int \rd^4 x \,{\rm d}^4\q {\rm d}^4{\bar \q}
\,E\, \frac{{ W}{\bar { W}}\cL^{++}}{({ \S}^{++})^2}~, 
\qquad E^{-1}= {\rm Ber}(E_A{}^M)~,
\label{InvarAc}
\eea
where
 \be
{ \S}^{++}:=\frac{1}{4}\Big( (\cD^+)^2 +4S^{++}\Big){ W}
=\frac{1}{4}\Big( ({\bar \cD}^+)^2 +4\bar{S}^{++}\Big){ {\bar W}}
=\S^{ij}u^+_i u^+_j~.
\label{Sigma}
\ee
Here the Lagrangian $\cL^{++}(z,u^+)$ is a covariant real projective 
multiplet of weight two, $ W(z)$ is the 
covariantly chiral field strength of an Abelian vector multiplet, 
$S^{++}(z,u^+)=S^{ij}(z)u^+_i u^+_j$ and 
$\bar{S}^{++}(z,u^+)={\bar S}^{ij}(z)u^+_i u^+_j$ 
are special dimension-1 components of the torsion.
The action (\ref{InvarAc}) can be shown to be invariant 
under the supergravity gauge transformations, and it is also manifestly 
super-Weyl invariant \cite{KLRT-M}.
It can also be rewritten in the equivalent form
\bea
S&=&
\frac{1}{2\pi} \oint_C (u^+ \rd u^{+})
\int \rd^4 x \,{\rm d}^4\q {\rm d}^4{\bar \q}
\,E\, \frac{\cL^{++}}{ S^{++} \bar{S}^{++}}~ 
\label{InvarAc2}
\eea
in which, however, the super-Weyl invariance is not manifest. 
The latter form makes transparent the fact that the action
is independent of the 
compensating vector multiplet described by $W$ and $\bar W$
provided $\cL^{++} $ is independent of it.

As argued in \cite{KLRT-M,K-2008}, the dynamics of a general $\cN=2$
supergravity-matter system can be described by an action of 
the form  (\ref{InvarAc}), including the chiral actions which can 
always be brought to the form  (\ref{InvarAc}).
This is why the action principle  (\ref{InvarAc}) 
is of fundamental importance in $\cN=2$ supergravity.

There are two special properties of the action (\ref{InvarAc}) that we would like to point out. 
{}First of all, the integration in   (\ref{InvarAc}) is carried out over the {\it full} superspace,
therefore one has to integrate out eight Grassmann variables in order to reduce 
the action to components. 
Secondly, the Lagrangian in (\ref{InvarAc}) obeys the analyticity  constraints 
(\ref{ana-introduction}) which enforce  $\cL^{++}$ 
to depend on only  {\it half} of the superspace Grassmann variables.
In this respect, the $\cN=2$ action (\ref{InvarAc}), 
or more precisely its equivalent form (\ref{InvarAc2}),  is analogous to the chiral action 
in 4D $\cN=1$ supergravity \cite{Zumino78,SG}, 
as specially emphasised in \cite{KTM-4D-confFlat}.
These two features of the $\cN=2$ supergravity action  hint at an opportunity 
to use the experience gained and the techniques developed, e.g.,  in 
4D $\cN=1$ superfield supergravity, 
in order to reduce (\ref{InvarAc}) to components.

In textbooks on 4D $\cN=1$ supergravity \cite{WB,GGRS,BK}, one can find 
two methods of component reduction. 
One of them (to be referred to as  {\it method 1}), 
elaborated in detail\footnote{More precisely,
Ref. \cite{GGRS} only stated the density formula and sketched its derivation. 
Years later, three of the authors of \cite{GGRS} came up with
simple  alternative derivations of the density formula \cite{Gates,GKS}.}
in \cite{WB,GGRS}, 
was originally introduced by Wess and Zumino 
\cite{WZ2} and presents itself as a version of the Noether procedure. 
It involves the following two steps: (i) starting from the superfield dynamical variables, 
one first reads off corresponding multiplets of component fields and their local 
supersymmetry transformations, using  a Wess-Zumino gauge imposed on the superfield
vielbein and connection; 
(ii) after that, the desired density multiplet 
is iteratively reconstructed from its lowest component 
in conjunction with the known supersymmetry 
transformation laws.
This method was further developed, and generalized to the case of chiral actions
in $\cN=2$ supergravity, in \cite{Muller82,Ramirez,Muller}
using covariant expansions with respect to  $\Theta$-variables \cite{WZ2,WB} 
of somewhat mysterious geometric origin.
The other approach ({\it method 2}) 
was elaborated in detail in \cite{BK}, although its first application in the case of pure 
supergravity was given by Gates and Siegel \cite{SG}.
It can be implemented provided there exists a formulation of the given supergravity theory in terms 
of unconstrained prepotentials, and such a formulation is indeed available in the case 
of 4D $\cN=1$ supergravity \cite{SG,OS}.
It involves the same step (i) as above modulo the fact that a Wess-Zumino gauge 
is now imposed on the supergravity prepotentials.
Its real gain is that, instead of carrying out the painfully laborious procedure (ii) 
of method 1, now one should simply do an ordinary 
Grassmann integral.

Both methods discussed above are hardly of any practical use in the case of 
$\cN=2$ supergravity formulation under consideration. 
Being applicable in principle, method 1 becomes too laborious 
to be used for general $\cN=2$ supergravity-matter systems.
As to method 2, no prepotential formulation is yet available for 
the projective-superspace formulation for $\cN=2$ supergravity given in \cite{KLRT-M}.
A prepotential formulation for $\cN=2$ supergravity  has been constructed within the 
harmonic-superspace approach \cite{GIKOS,GIOS,GIOS-book}.\footnote{In 
the rigid supersymmetric case, the harmonic \cite{GIKOS} and 
the projective \cite{KLR,LR} approaches are closely related  \cite{K-double}, 
and this should extend, in principle, to the case of supergravity.} However, no comprehensive 
analysis of the component reduction in curved harmonic superspace has yet appeared.

A relatively new paradigm for component reduction in supergravity
  appeared some ten years ago.
As advocated in Refs.  \cite{GKS,GK}, which built on the earlier work \cite{AD}, 
an ideal means to perform covariant 
theta-expansions and integrate out Grassmann variables is provided by 
the superspace normal coordinates introduced a quarter of a century ago by McArthur \cite{McA}
for completely different aims.\footnote{In \cite{McA2}, the normal coordinate techniques
\cite{McA} were applied
to compute the so-called $b_4$ (or, equivalently,  $a_2$) coefficients for chiral matter in 
4D $\cN=1$ supergravity. Although there exists a purely covariant and very efficient 
approach to evaluate the Schwinger-DeWitt coefficients in curved superspace  \cite{BK86}, 
the method of superspace normal coordinates \cite{McA} 
proves to be  truly indispensable for deriving 
the density formulae in supergravity theories, as emphasized in  \cite{GKS}.}
This technique was applied in  \cite{GKS,GK} to compute the density formula 
for several supergravity models in diverse dimensions including the case of 4D 
$\cN=1$ supergravity.
Since the method of fermionic normal coordinates employed in
 \cite{GKS,GK} is a version of Wess-Zumino gauge in curved superspace, 
this construction is ultimately related to the earlier approaches pursued in 
 \cite{Muller82,Ramirez,Muller}.

The powerful property  of the method of normal coordinates\footnote{In 
$\cN=1$ supergravity, there exists a different normal coordinate construction 
\cite{OS-normal} based on the prepotential formulation due to Ogievetesky and Sokatchev 
\cite{OS}. This normal gauge should possess a natural extension to the case 
of $\cN=2$ supergravity formulated in harmonic superspace \cite{GIKOS,GIOS,GIOS-book}, 
and it would be very interesting to work out such an extension explicitly. } 
 \cite{McA} is its universality,
as emphasized in \cite{GKS} (of course, this is not accidental, 
for the method is a superspace extension of the Riemann normal coordinates).
It can be used for any supergravity theory formulated in superspace, 
for any number of space-time dimensions. For example, it has recently been used
in the case of eleven dimensional supergravity \cite{Ts}.
In particular, it can be applied to reduce the action (\ref{InvarAc}) 
to components. However, the latter application would still require a nontrivial 
computational  effort. Remarkably, the specific feature of 4D $\cN=2$ supergravity
(and also 5D $\cN=1$ supergravity) 
is that it offers us an alternative and much more  efficient scheme  
to reduce the action (\ref{InvarAc}) to components which is based on  the principle of 
projective  invariance \cite{K-hyper1,KT-M,KT-Msugra1}. 
This unusual invariance, which has no analogue in the $\cN=1$ case, 
is easy to visualize in  a flat superspace limit where the action (\ref{InvarAc}) reduces to 
\bea
S_{\rm flat}&=&\frac{8}{\pi} \oint (u^+ \rd u^{+})
\int \rd^4 x \,{\rm d}^4\q {\rm d}^4{\bar \q}
\, \frac{W{\bar W}L^{++}(u^+)}{(D^+)^2W\,(\DB^+)^2\bar{W}}
\non \\
&=& \frac{1}{2\pi}  \oint \frac{(u^+\rd u^{+})}{(u^+u^-)^4}
\int \rd^4 x \, (D^-)^2(\DB^-)^2 L^{++}(u^+)\big|_{\q = \bar \q =0}~.
\label{flatac}
\eea
Here the spinor derivatives $D^-_\a$ and ${\bar D}^-_\ad$ 
are obtained from $D^+_\a$ and ${\bar D}^+_\ad$ by replacing 
$u^+_i \to u^-_i$, with the latter being a fixed constant 
isotwistor for which  the only constraint is
$(u^+u^-)\neq 0$ at each point of  the integration contour. 
Since $L^{++}$ is a weight-two rigid projective supermultiplet, 
the action can be seen to be invariant 
under arbitrary {\it projective transformations} of the form:
\be
(u_i{}^-\,,\,u_i{}^+)~\to~(u_i{}^-\,,\, u_i{}^+ )\,R~,~~~~~~R\,=\,
\left(\begin{array}{cc}a~&0\\ b~&c~\end{array}\right)\,\in\,{\rm GL(2,\mathbb{C})}~.
\label{projectiveGaugeVar}
\ee
Clearly, this projective invariance is almost obvious in flat superspace.
In curved superspace, however, it turns into a powerful constructive principle 
to reduce  the action (\ref{InvarAc}) to components, and what is most
non-trivial -- without imposing any Wess-Zumino  gauge condition!

This paper is organized as follows. In section 2, we provide an alternative
derivation of normal coordinates around a submanifold in an arbitrary  curved superspace. 
Although the consideration given in \cite{GKS} involves some ingenious acrobatics, 
it leaves several important questions unanswered such as  
the explicit structure of equations which could allow one to derive 
normal coordinate expressions  
for the connection and the vielbein to any order in perturbation theory
(in this respect, the work \cite{Ts}, which closely follows the original 
normal coordinate construction of \cite{McA}, contains very useful results).
Our presentation in section 2 is based in part on earlier approaches developed
in general relativity \cite{Synge} and quantum gravity 
\cite{DeWitt1,DeWitt2,BV,Avr} many years ago, as well as some more 
recent covariant techniques for super Yang-Mills theories \cite{KMcA}.\footnote{The
material in section 2 is based in part on unpublished lecture notes by one of us (SMK)
\cite{K-Hannover}.}
Here we  derive differential equations which are obeyed by the vielbein and the connection 
in normal coordinates,
and which can be used to reconstruct these objects, in principle in closed form. 
We also present an equation for the super-determinant of the vielbein,
 $E= {\rm Ber} (E_M{}^A)$, 
which allows one to reconstruct $E$ without a detailed knowledge of the vielbein.
As an application of the techniques developed in section 2, 
in section 3 we explicitly  reduce
an integral over the full 4D $\cN=2$ curved superspace to that over the chiral subspace.

Section 4 is central to the present work. Here we reduce the action (\ref{InvarAc})
to components using the principle of projective invariance. 
We also consider two applications. First, we prove the gauge invariance of the 
special vector-tensor coupling introduced in \cite{KLRT-M}. 
Second, we give a curved superspace description for the c-map \cite{cmap1,cmap2}.
In section 5, we derive a new representation for the covariantly chiral projector 
and use this result to reformulate the action (\ref{InvarAc}) as a chiral integral.

This paper is accompanied by three technical appendices. 
In appendix A we collect the salient points of the superspace formulation for 
$\cN=2$ supergravity, following \cite{KLRT-M}, which are essential for understanding 
the main results of this paper.  Appendix B summarizes the main properties of 
covariant projective supermultiplets following \cite{KLRT-M}. Finally, 
appendix C provides the proof of eq. (\ref{chiralproj2}).

\section{Integrating out fermionic dimensions }
\setcounter{equation}{0}

In this section, we temporarily leave aside the main object of our study -- $\cN=2$
matter-coupled supergravity in four space-time dimensions, and instead discuss
the problem of defining a normal  coordinate system around a submanifold of 
a curved superspace with any number of bosonic and fermionic dimensions.
We will present an application of the formalism developed to the case of 4D $\cN=2$ 
supergravity in section 3.

\subsection{Parallel transport and associated two-point functions}
\label{parallel1}
Let us consider a curved superspace $\cM \equiv \cM^{d|\d}$ with $d$ space-time 
and $\d$ fermionic dimensions, and  let $z^M$ be local coordinates chosen to parametrize  $\cM$.
The corresponding superspace geometry is described by covariant derivatives 
\bea
\cD_A =E_A +\F_A~, \qquad 
E_A : = E_A{}^M(z)\, \pa_M ~, \qquad 
\F_A := \F_A (z) {\bm \cdot} {\mathbb J} =E_A{}^M \F_M ~.
\eea
Here $\mathbb J$ denotes 
the generators of the structure group\footnote{The formalism below can be readily generalized 
to incorporate an internal Yang-Mills group $G_{\rm int}$ by replacing
$G \to G \times  G_{\rm int}$.}
 $G$ (with all indices of $\mathbb J$s  suppressed),  
$E_A $ is the inverse vielbein, and $\F ={\rm d} z^M \F_M =E^A \F_A$ the connection. 
As usual, the matrices defining the vielbein $E^A := {\rm d}z^M E_M{}^A(z)$
and its inverse  $E_A $ obey  the identities $E_A{}^M E_M{}^B =\d_A{}^B$ and
$E_M{}^A E_A{}^N =\d_M{}^N$. 
An infinitesimal $G$-transformation acts on the components of a vector field
$v =v^A E_A$ and a one-form $\o= E^A \o_A$ as follows:
\bea
[\l {\bm \cdot} {\mathbb J} ,  v^A] = \l^A{}_B v^B = -v^B \l_B{}^A
~, \qquad 
[\l {\bm \cdot} {\mathbb J} ,  \o_A] = - \o_B \l^B{}_A  =  \l_A{}^B \o_B~,
 \eea
such that $(v) \o := v^A \o_A$ is invariant. Here we 
have  assumed that the structure group transformations preserve the 
Grassmann parity $\ve$ of any tensor superfield, which requires $\ve(\l_A{}^B) =0$,
and the transformation parameters  are defined to obey  $\l_A{}^B = - \l^B{}_A$.

The covariant derivatives obey the algebra 
\bea
[\cD_A , \cD_B \} = T_{AB}{}^C \cD_C + R_{AB}{\bm \cdot} {\mathbb J} ~, 
\eea
with $T_{AB}{}^C $ the torsion, and $R_{AB}$ the curvature of $\cM$.
In particular, 
\bea
\{\cD_A,\cD_B\}\,\o_C=T_{AB}{}^D\cD_D\,\o_C
+R_{AB}{}_C{}^D\, \o_D~,
\label{torsion-curvature}
\eea
when acting on the one-form $\o_A$.

It is pertinent to our consideration to recall the basic facts about parallel transport. 
Let $z' \in \cM$ be a given superspace point, and $\g (t)  = \{z^M(t)\}$ a smooth curve
in $\cM$ such that $\g(0) =z'$. For the tangent vector to $\g$ at $z(t)$, we convert 
its world index into a local flat one, 
\bea
\z^A(t) := \dt{z}{}^M(t) E_M{}^A(z(t)) ~.
\label{zeta}
\eea
Let $v^{A'}= v^{M'} E_{M'}{}^{A'}(z')$ be a tangent vector at $z'$, $v \in T_{z'}\cM$.
Its parallel transport along $\g$, $v(t) \in T_{z(t)}\cM$,  is defined to satisfy the equation
\bea
\Big( \frac{\rm d}{{\rm d} t} + {\z}^B (t) \F_B (t) \Big)  v^A(t) 
=0~.
\eea
The parallel transport of a tensor $\cV'$ at $z'$ along  the curve $\g (t)$ is defined similarly. 
 
All information about parallel transport along the curve $\g (t)$ is encoded in the 
corresponding  {\it parallel displacement propagator along} $\g$, $I_\g (t) \in G$,
which is defined by the following conditions:\\
${}\qquad$(i) the parallel transport equation
\bea
\Big( \frac{\rm d}{{\rm d} t} + {\z}^B (t) \F_B (t) \Big) I_\g(t)=0~;
\label{pt-equation}
\eea
 ${}\qquad$(ii) the initial condition
 \bea
 I_\g(0) = \mathbbm{1}~.
 \label{pt-initial}
 \eea
Then, for any tensor $\cV'$ at $z'$, its parallel transport along $\g (t)$ is 
\bea
\cV(t) = D\Big( I_\g (t) \Big) \cV'~,
\eea
where $D$ is the representation of the structure group $G$ in which  the tensor 
transforms.\footnote{In what follows, we do not indicate explicitly the representation $D$
of the structure group,
and the matrix $D\big( I_\g (t) \big)$ will always be written simply as $ I_\g (t) $.}
As is known, a unique solution to eqs. (\ref{pt-equation}) and 
(\ref{pt-initial}) is the path-ordered exponential
\bea
I_\g(t) = {\rm P} \,{\rm e}^{-\int_\g \F}~.
\eea
The important feature of the equation (\ref{pt-equation}) is its invariance 
under reparametrizations of the curve.

Now, let $\hat{\g} (t)= \{z^M(t)\}$ be a geodesic through
$z'$,
\bea
\Big( \frac{\rm d}{{\rm d} t} + {\z}^B (t) \F_B (t) \Big) \z^A (t) =0~, 
\qquad \hat{\g}(0) =z'~.
\label{geodesicequation}
\eea
For any point $z^M(t)$ on the geodesic, we define $I\big(z(t); z'\big) := I_{\hat{\g}}(t)$.
Since any two points $z'$ and $z$ in $\cM$ can be connected by a geodesic, 
which is locally unique modulo worldline reparametrizations, 
we obtain a well-defined two-point function 
\be
I(z;z') \in G ~, \qquad  I(z'; z') = \mathbbm{1}~.
\ee
It will be called the parallel displacement propagator.

The freedom to choose affine parametrization of the geodesic, which connects
$z'$ and $z$, can be fixed as
\bea
z' = \hat{\g} (0)~, \qquad z = \hat{\g}(1)~,
\eea
which corresponds to the standard exponential mapping (see, e.g., \cite{Willmore}).
{}For this parametrization, we define vector two-point functions\footnote{In the case 
when $\cM$ is an ordinary  Riemannian manifold, in particular if $T_{AB}{}^C =0$,
one can show that $\z^A (z,z')=\cD^A \s (z,z') $ and   $\z^{A'}(z';z) = \cD^{A'} \s(z,z')$, 
where $\s(z,z') =\s(z',z)$ is the so-called {\it world function} coinciding with half the square 
of the geodesic distance between the points $z'$ and $z$, see  
\cite{Synge,DeWitt1,DeWitt2} for more detail.
In the mathematics literature,  the $\s(z,z')$ is sometimes 
referred  to as the {\it distance function} \cite{Willmore}.}
\begin{subequations}
\bea
\z^A(z;z') &:=& \z^A(t=1) \in T_z\cM~,
\label{zeta-tpf1}
\\
\z^{A'}(z';z) &:=& -\z^A(t=0) \in T_{z'}\cM~.
\label{zeta-tpf2}
\eea
\end{subequations}
These functions are related to each other as 
follows:
\bea
\z^{A}(z;z') = -\big[ I(z;z')\big]^A{}_{B'} \,\z^{B'}(z';z)~.
\label{vector-connect}
\eea
The parallel displacement propagator, $I(z;z')$, obeys the differential equations:
\begin{subequations}
\bea
\z^B \cD_B I(z;z') &=&0~,
\label{pdp-eq1}
\\
\z^{B'} \cD_{B'} I(z;z') &=&0~.
\label{pdp-eq2}
\eea
\end{subequations}
These equations follow from (\ref{pt-equation}). 
It also holds that 
\be
I(z;z') \,I(z';z) = \mathbbm{1}~.
\ee
As to the two-point functions $\z^A(z;z') $ and $\z^{A'}(z';z)$, 
they enjoy the following equations:
\begin{subequations}
\bea
\z^B \cD_B \z^A &=&\z^A~,
\label{vector-tpf-1}
\\
\z^B \cD_B \z^{A'} &=& \z^{A'}~.
\label{vector-tpf-2}
\eea
\end{subequations}
To prove eq. (\ref{vector-tpf-1}), 
it suffices to note that for a geodesic $z^M(t)$ 
passing through $z'$,  $z(0) =z'$, we have 
\bea
\z^A(z(t); z') &=& t \,\z^A(t)~, 
\eea
with $\z^A(t)$ the tangent vector to the given geodesic at $z(t)$. 
Then, it only remains to use the geodesic equation
(\ref{geodesicequation}). As to equation (\ref{vector-tpf-2}), 
it now follows from the relations (\ref{vector-connect}), (\ref{pdp-eq1})
and (\ref{vector-tpf-1}).

\subsection{Covariant Taylor expansion}
\label{Taylor}

Let $\cV(z)$ be a tensor superfield transforming in some representation 
of the structure group. Then it can be expanded in a covariant Taylor series 
of the form: 
\bea
I(z';z)  \cV(z) = 
\sum_{n=0}^{\infty} \frac{(-1)^n}{n!} \z^{A'_n} \dots \z^{A'_2} \z^{A'_1} \,
\cD_{A'_1} \cD_{A'_2} \dots \cD_{A'_n} \cV(z')~.
\label{Taylor1}
\eea
It can be justified simply  by generalizing the proof given, e.g.,  in \cite{BV}
for the case when $\cM$ is a Riemannian manifold.

\subsection{Parallel transport around the submanifold}
\label{parallel2}

Up to now, we have considered all possible geodesics 
passing through a fixed point $z' \in \cM$, 
where the latter have been  completely arbitrary.
${}$From now on, we turn to a more general setup.  First of all, 
we will restrict $z'$ to belong 
to a fixed submanifold $\S\equiv \S^{d'|\d'}$ of the superspace $\cM= \cM^{d|\d}$, 
with $\d'<\d$ or/and $d'<d$. Secondly, 
we will only consider those geodesics $\hat{\g}(t)$ 
through $z'$, $\hat{\g}(0)=z'$, which are 
transverse to $\S$.  To make the latter requirement more precise, we assume 
in addition that the vielbein $E^A$s can be split into two disjoint subsets, 
\be
E^A = ( E^{\hat{a}}, E^{\hat{\a}} )~, 
\ee
such that the set of one-forms $E^{\hat{a}}|_{z'}$ constitutes a basis 
of the cotangent space  $T^*_{z'}\S$ at any point $z' \in \S$. 
Then, the requirement that $\hat{\g}(t)$ be transverse to $\S$, 
will mean the following: 
\be
\dt{z}{}^M (0) \,E_M{}^{\hat a}(z')
=0~, \qquad z(0) =z' \in \S~.
\label{normal}
\ee
${}$Finally, we put forward one more technical requirement, that the structure 
group $G$ acts reducibly on $E^A$s such that each of 
the two subsets $E^{\hat{a}}$s ad $E^{\hat{\a}} $s
transforms into itself under the action of $G$.
The setup introduced here reduces to that considered in subsection \ref{parallel1} 
if $\S$  shrinks down to  a single point $z'$.

Let $\tilde{z}^{\hat m}$ be local coordinates parametrizing  
the submanifold $\S$. These variables can be extended 
to provide a local coordinate system $z^M = (\tilde{z}^{\hat m}, {y}^{\hat \m})$
in  the whole superspace $\cM$ in such a way 
that along $\S$ we have 
\be
z^M\big|_\S =  (\tilde{z}^{\hat m}, {y}^{\hat \m}=0)~.
\ee
Reparametrization invariance can be further used to choose
\bea
E_M{}^A(z) \big|_\S = \left(
\begin{array}{cc}
\cE_{\hat m}{}^{\hat a}(\tilde{z})   ~ & \cE_{\hat m}{}^{\hat \a}(\tilde{z})   \\
0     ~ & \d_{\hat \m}{}^{\hat \a}  
\end{array}
\right) ~.
\label{coordinategauge}
\eea
Then, eq. (\ref{normal}) becomes 
\bea
\dt{z}{}^M (0) = \big(0, \dt{y}{}^{\hat \m} (0) \big)~.
 \eea
In terms of $\z^A(t)$, eq. (\ref{zeta}),  this is equivalent to 
\bea
\z^A(0) = \z^{\hat \m} \d_{\hat \m}{}^A~, \qquad 
\z^{\hat \m} \equiv \dt{y}^{\hat \m} (0) ~.
\label{der}
\eea
It follows from the above consideration that 
\bea
\z^{\hat a}(z;z') = \z^{\hat{a}{}'}(z';z) = 0~.
\eea

As an example, let us consider a curved superspace 
corresponding to  four-dimensional $\cN=2$ conformal supergravity reviewed in 
Appendix \ref{SCG}. It follows from the anticommutation relations (\ref{acr2})
that the vector fields\footnote{The inverse vielbein is thus $E_A =(E_{\hat a},E_{\hat \a})$,
where $ E_{\hat a} := (E_a, E_\a^i)$.}
$E_{\hat \a}:= {\bar E}^\ad_i$ generate an involutive distribution (see, e.g., 
\cite{Willmore} for a review of the relevant differential-geometric constructions), that is
\bea
\{ {\bar E}^\ad_i ,{\bar E}^\bd_j \} = C^\ad_i{\,}^\bd_j{\,}_\gd^k (z) {\bar E}^\gd_k~.
\eea
Then, the Frobenius theorem (see, e.g., \cite{Willmore}) 
implies that one can replace the original local coordinates $z^M$ 
by new ones,
$\{\tilde{z}^{\hat m}, {\r}^{\hat \m}\}$, with the properties:
\bea
E_{\hat \a} \tilde{z}^{\hat m}=0~,\qquad
E_{\hat \a} =N_{\hat \a}{}^{\hat \m} (\tilde{z}, \r) \frac{\pa}{ \pa \r^{\hat \m} }~, 
\label{eq-chiral-1}
\eea
for some non-singular matrix $N_{\hat \a}{}^{\hat \m}$. It is clear that covariantly chiral 
scalar superfields, ${\bar \cD}^\ad_i \F=0$, are functions of the variables 
$\tilde{z}^{\hat m} $, 
$\F =\F(\tilde{z})$.
The submanifold $\S$ in the above discussion will be identified
 with the chiral subspace defined by the equations $\r^{\hat \m}=0$.
Replacing $\r^{\hat \m}$ by  new variables $y^{\hat \m} $ defined as
\be
\r^{\hat \m} = y^{\hat \n}\, \d_{\hat \n}{}^{\hat \a} \,N_{\hat \a}{}^{\hat \m}(\tilde{z}, \r)~,
\label{eq-chiral-2}
\ee
one can see that the inverse vielbein  restricted to $\S$ has the form:
\bea
E_A{}^M (z) \big|_\S = \left(
\begin{array}{cc}
\cE_{\hat a}{}^{\hat m}(\tilde{z})   ~ & \cE_{\hat a}{}^{\hat \m}(\tilde{z})   \\
0     ~ & \d_{\hat \a}{}^{\hat \m}  
\end{array}
\right) ~.
\label{coordinategauge2}
\eea
This result is equivalent to (\ref{coordinategauge}).
In the example considered, the involutive distribution 
generated by ${\bar E}^\ad_i$, determines all the tangent vectors 
being transverse to $\S$.

\subsection{Normal coordinates around the submanifold}
\label{Normal-around}

A normal coordinate system\footnote{In  Riemannian geometry, 
normal coordinates around a submanifold were discussed in \cite{Synge2}.}
 around $\S$ is defined by the following two conditions:

(i) All 
geodesics, which are transverse to $\S$, are straight lines.
\bea
\tilde{z}^{\hat m} (t) = \tilde{z}^{\hat m} ~, \qquad 
{y}^{\hat \m} (t) = t\, \z^{\hat \m} ~.
\label{NG1}
\eea
Such a geodesic connects the superspace points $(\tilde{z}, 0) $ and  $(\tilde{z}, \z) $.

(ii) Fock-Schwinger (or structure group) gauge: 
\bea
I\big(z;z'\big) =I\big( \tilde{z}, \z ; \tilde{z}, 0\big) ={\mathbbm 1}~.
\label{NG2}
\eea

${}$For the two-point function $\z^A(z,z')$, eq.  (\ref{zeta-tpf1}), 
the condition (\ref{NG1}) implies 
\bea
\z^A(z;z') = \z^{\hat \m} E_{\hat \m}{}^A(\tilde{z}, \z) \equiv
\z^M E_M{}^A(\tilde{z},\z) ~, \qquad 
\z^M:= (0,  \z^{\hat \m} )~.
\eea
${}$For the two-point function $\z^{A'}(z';z)$, eq.  (\ref{zeta-tpf2}), 
the condition (\ref{der}) gives
\bea
\z^{A'}(z';z) = -\z^M \d_M{}^A~.
\eea
Now, using eqs. (\ref{vector-connect}) and (\ref{NG2})  gives
\bea
\z^M E_M{}^A(\tilde{z}, \z)   = \z^M \d_M{}^A=\z^{\hat \m} \d_{\hat \m}{}^{\hat \a}~.
\label{NG3}
\eea
${}$Furthermore, using eqs. 
(\ref{pdp-eq1})
and  (\ref{NG2})  gives
\bea
\z^A \F_A(\tilde{z},\z) {\bm \cdot} {\mathbb J}= \z^M \F_M (\tilde{z}, \z) {\bm \cdot} {\mathbb J}=0~.
\label{NG4}
\eea
The relations (\ref{NG3}) and (\ref{NG4}) are the key results for applications.\footnote{In
the zero-dimensional case when $\S$ reduces to a single point $z'$, 
the relations (\ref{NG3}) and (\ref{NG4}) are equivalent to those given in \cite{McA}.
In the case when $\S= \S^{(d,0)}$ is the bosonic body of the curved superspace 
$\cM= \cM^{(d,\d)}$,  the relations (\ref{NG3}) and (\ref{NG4}) were derived in 
\cite{Ts} in a different manner.} These relations did not appear in \cite{GKS}.
It is worth pointing out that eq. (\ref{NG4})  implies 
\be
\F_{\hat \a}(\tilde{z},0) {\bm \cdot} {\mathbb J}
= \F_{\hat \m} (\tilde{z}, 0) {\bm \cdot} {\mathbb J}=0~, 
\ee
while no restriction is imposed on $\F_{\hat m} (\tilde{z}, 0) {\bm \cdot} {\mathbb J}$ 
which is the connection on $\S$.

Relations (\ref{NG3}) and (\ref{NG4}) can be rewritten in terms of the operation 
of interior product, $\imath_\z$. It is worth recalling how the latter is defined.
Given a vector field $\cV = \cV^M \pa_M = \cV^A E_A$ and a $p$-form
\be
\O = \frac{1}{p!} {\rm d} z^{M_p} \dots {\rm d} z^{M_1} \O_{M_1 \dots M_p}
=\frac{1}{p!} E^{A_p} \dots E^{A_1} \O_{A_1 \dots A_p}~,
\ee
the $(p-1)$-form $\imath_\cV \O$ is defined as
\be
\imath_\cV \O
= \frac{1}{(p-1)!} {\rm d} z^{M_p} \dots {\rm d} z^{M_2}  \cV^{M_1} \O_{M_1 \dots M_p}
=\frac{1}{(p-1)!} E^{A_p} \dots E^{A_2} \cV^{A_1}\O_{A_1 \dots A_p}~.
\ee
Now, eqs. (\ref{NG3}) and (\ref{NG4}) can be rewritten as follows:
\begin{subequations}
\bea
\imath_\z E^A &=& \z^M \d_M{}^A=\z^{\hat \m} \d_{\hat \m}{}^{\hat \a}~,
\label{NG3-inter} \\
\imath_\z \F_A{}^B &=&0~,
\label{NG4-inter}
\eea
\end{subequations}
 with $ \F_A{}^B  = {\rm d} z^{M} \F_{MA}{}^B = E^C \F_{CA}{}^B$ the connection
 one-form.
 
\subsection{Structure equations}
We turn to uncovering the implications of eqs. (\ref{NG3-inter}) and 
 (\ref{NG4-inter}), building on the construction in Riemannian geometry 
given in \cite{ABO,MSvdV}.

We start by introducing the torsion two-form
\bea
T^A = \hf E^CE^B T_{BC}{}^A
\eea
and the curvature two-form
\bea
R{\bm \cdot} {\mathbb J} = \hf E^D E^C R_{CD}{\bm \cdot} {\mathbb J}~, \qquad 
[R{\bm \cdot} {\mathbb J} , \o_A ] = R_A{}^B \o_B
=\hf E^D E^C R_{CDA}{}^B \o_B~,~~~
\eea
with $\o_A$ an arbitrary one-form.
They obey the structure equations:
\begin{subequations}
\bea
-T^A &=& {\rm d} E^A -E^B \F_B{}^A~,
\label{str-eq1} \\
R_A{}^B&=& {\rm d} \F_A{}^B -\F_A{}^C \F_C{}^B~.
\label{str-eq2} 
\eea
\end{subequations}

Let us make use of the well-known differential geometric relation
\be
L_\z = \imath_\z\, {\rm d} + {\rm d}\,\imath_\z~,
\label{Lie}
\ee
with $L_\z $ the Lie derivative.
Applying both sides of 
this relation to $\F_A{}^B$ and 
using the structure equation (\ref{str-eq2}) and the gauge condition
(\ref{NG4-inter}), we obtain
\bea
L_\z \F_A{}^B = \imath_\z R_A{}^B~.
\label{str-connection1}
\eea
Similarly we can evaluate $L_\z E^A$ to obatin 
\bea
L_\z E^A = \cD \z^A -\imath_\z T^A~, 
\qquad 
\cD \z^A :={\rm d}\z^A - \z^B \F_B{}^A~.
\label{str-curvature1}
\eea
Applying again $L_\z$ to both sides of (\ref{str-curvature1}) and making use 
of the gauge conditions and the structure equations, one obtains
\bea
(L_\z -1) L_\z E^A &=& -\cD \z^D \,\z^C T_{CD}{}^A 
+ (\imath_\z T^D)\, \z^C T_{CD}{}^A
-E^D \z^C 
L_\z T_{CD}{}^A \non \\
&& - \z^B \imath_\z R_B{}^A~.
\label{str-curvature2}
\eea
Here the Lie derivative of the torsion tensor can be represented, 
due to (\ref{NG4}), as
\bea
L_\z T_{CD}{}^A = \z^{\hat \n} \pa_{\hat \n} T_{CD}{}^A
=\z^{\hat \b} \cD_{\hat \b}T_{CD}{}^A ~.
\eea

The Lie derivative of a one-form is 
\bea
L_\cV \o_M = \cV^N \pa_N \o_M +\big( \frac{\pa}{\pa z^M} \cV^N\big) \o_N~,
\label{Lie2}
\eea
and thus
\bea
L_\z \o_M = \z^{\hat \n} \pa_{\hat \n} \o_M +\d_M{}^{\hat \n} \o_{\hat \n}
\quad \Longrightarrow \quad 
 \left\{
\begin{array}{l}
L_\z \o_{\hat m} =\z\cdot \pa\, \o_{\hat m}  \\
 L_\z \o_{\hat \m} =(\z\cdot \pa +1) \o_{\hat \m}  ~.
\end{array}
\right. 
\label{Lie3}
\eea

The relations (\ref{str-connection1}) and (\ref{str-curvature1}), 
and their  corollary (\ref{str-curvature2}),  allow us to reconstruct 
the connection $\F_{MA}{}^B(\tilde{z}, \z)$ and the vielbein 
$E_{M}{}^A(\tilde{z}, \z)$ as Taylor series in $\z$s, in which all the coefficients
(except the leading $\z$-independent terms) 
are tensor functions of the torsion, the curvature and their covariant derivatives 
evaluated at  $ \z=0$ (of course, there also occur contributions involving the field
$ \cE_{\hat m}{}^{\hat \a}(\tilde{z}) $ defined in (\ref{coordinategauge})).
Indeed, consider a tensor superfield $\cV$ such as
$T_{CD}{}^A$ or $R_{CDB}{}^A$ and their covariant derivatives.
In the normal gauge, the covariant Taylor expansion, eq. (\ref{Taylor1}), becomes
\bea
\cV(\tilde{z}, \z) &=& 
\sum_{n=0}^{\infty} \frac{1}{n!} \z^{{\hat \a}_n} \dots  \z^{{\hat \a}_1} \,
\cD_{{\hat \a}_1}  \dots \cD_{{\hat \a}_n} \cV(\tilde{z},0)
\equiv \sum_{n=0}^{\infty} \cV^{(n)}~,
\quad  \z\cdot \pa \cV^{(n)} =n  \cV^{(n)}~,~~~~~~
\label{Taylor2}
\eea
with $\z^{\hat \a} \equiv \z^{\hat \m} \d_{\hat \m}{}^{\hat \a}$.
Eq. (\ref{str-connection1}) can be rewritten in the component form: 
\bea
L_\z \F_{MA}{}^B (\tilde{z}, \z)=  E_M{}^D (\tilde{z}, \z) \,\z^{\hat \g} \,
R_{{\hat \g} D A}{}^B (\tilde{z}, \z)~,
\label{str-connection2}
\eea
and similarly for eq. (\ref{str-curvature1}) 
or its corollary (\ref{str-curvature2}). Now, all tensors involved 
have to be represented by covariant 
Taylor series of the form (\ref{Taylor2}), while  $\F_{MA}{}^B(\tilde{z}, \z)$ and 
$E_{M}{}^A(\tilde{z}, \z)$  have to be given as ordinary Taylor series, in 
particular 
\bea
E_M{}^A (\tilde{z}, \z)= 
\sum_{n=0}^{\infty} \frac{1}{n!} \z^{ {\hat \n}_n} \dots  \z^{ {\hat \n}_1} \,
\pa_{{\hat \n}_1}  \dots \pa_{{\hat \n}_n} E_{ M }{}^A (\tilde{z},0)
\equiv \sum_{n=0}^{\infty} E^{(n)}{}_{ M }{}^A
~.
\label{Taylor3}
\eea
In accordance with (\ref{Lie3}), the Lie derivative $L_\z$ acts  on 
$E^{(n)}{}_{ M }{}^A$ in (\ref{Taylor3}), 
which is homogeneous of $n$-th degree in $\z$,  
as the operator of multiplication by $n$ if $M=\hat m$ or by $(n+1)$ if $M=\hat \m$.

\subsection{Computing the determinant of the vielbein}

Of crucial importance is  the explicit $\z$-dependence 
of the determinant $E:= {\rm Ber} (E_M{}^A)$.
The simplest way to address this problem is to derive a differential equation 
obeyed by  $E$ that follows from the equations given in the previous section. 

Using the standard identity $\d E = (-1)^M E \, \d E_M{}^A \, E_A{}^M$ in conjunction 
with eq. (\ref{Lie3}), we  obtain 
\bea
\z\cdot \pa \ln E = (-1)^M \big[ L_\z E_M{}^A 
- \d_M{}^{\hat \n} E_{\hat \n}{}^A \big] E_A{}^M~.
\eea
The right-hand side here can be transformed using the structure equation 
(\ref{str-curvature1}) to get 
\bea
\z\cdot \pa \ln E = -(-1)^A \big[ \F_{A {\hat \b} }{}^A \z^{\hat \b} 
+  \z^{\hat \b} T_{\hat \b A}{}^A \big]
+(-1)^{\hat \m} \d_{\hat \m}{}^{\hat \a}
\big(E_{\hat \a}{}^{\hat \m} - \d_{\hat \a}{}^{\hat \m}\big)~.
\label{E-master}
\eea
This is the master equation to determine the $\z$-dependence 
of $E =E(\tilde{z}, \z)$ under the boundary condition 
$E(\tilde{z}, 0) = \cE(\tilde{z}) $, where $\cE =   {\rm Ber}
\big(\cE_{\hat m}{}^{\hat a}   \big)$ is the determinant of the vielbein 
on the submanifold, as introduced in  eq. (\ref{coordinategauge}).
Eq. (\ref{E-master})  shows that one has to know the  $\z$-dependence 
of the connection in order to evaluate that of $E$. This result is quite nice and, 
at the same time, somewhat counter-intuitive, 
for one usually evaluates the vielbein only, while 
the explicit structure of the connection is completely  ignored. For instance, 
the authors of \cite{GKS} use a more laborious approach, which is:  
(i) to compute the $\z$-dependence of the vielbein $E_M{}^A$
by iterations; and then 
(ii) to evaluate  the determinant of the vielbein. 

Equation (\ref{E-master}) can be rewritten in a  somewhat different form 
if one recalls that the structure group has been assumed to act reducibly, 
that is $ \F_{A {\hat \b} }{}^C = \F_{A {\hat \b} }{}^{\hat \g}\, \d_{\hat \g}{}^C$.
This gives
 \bea
\z\cdot \pa \ln E = -(-1)^{\hat \a} \F_{{\hat \a} {\hat \b} }{}^{\hat \a} \z^{\hat \b} 
-(-1)^A  \z^{\hat \b} T_{\hat \b A}{}^A 
+(-1)^{\hat \m} \d_{\hat \m}{}^{\hat \a}
\big(E_{\hat \a}{}^{\hat \m} - \d_{\hat \a}{}^{\hat \m}\big)~.
\label{E-master2}
\eea
It often happens that 
\bea
(-1)^A  T_{\hat \b A}{}^A =0~.
\eea
In particular, such a  situation occurs in  the cases of $\cN=1$ and $\cN=2$ supergravity 
when $\z^{\hat \a}$ are Grassmann coordinates.
In this case we end up with the remarkably simple equation: 
 \bea
\z\cdot \pa \ln E = -(-1)^{\hat \a} \F_{{\hat \a} {\hat \b} }{}^{\hat \a} \z^{\hat \b}  
+(-1)^{\hat \m} \d_{\hat \m}{}^{\hat \a}
\big(E_{\hat \a}{}^{\hat \m} - \d_{\hat \a}{}^{\hat \m}\big)~.
\label{E-master3}
\eea

%%%%%%%%%%%
%%%%%%%%%%%%%%%%%%%%%%%%%%

\section{Reduction to chiral subspace in $\cN=2$ supergravity }
\setcounter{equation}{0}

As an illustration of 
the normal coordinate techniques developed in 
section 2,  here we apply the scheme to the case when  
$\cM$ is the curved 4D $\cN=2$ superspace as defined in Appendix A, 
and $\S$ its chiral subspace. 
All the relevant information regarding the chiral subspace 
can be found at the end of subsection 2.3.
Our goal is to reduce an integral over the full superspace,  
$\int \rd^4 x \,{\rm d}^4\q{\rm d}^4{\bar \q}\,E\, U$, 
to that over the chiral subspace, for any scalar and isoscalar superfield $U$.

In this section we  continue to use the ``hat'' index notation, which was introduced in 
 section 2, as much as possible, keeping in mind that, 
 for instance,  $\cD_{\hat \a}:= {\bar \cD}^\ad_i$. 
We also use  the notation  (\ref{Taylor2}), 
with $  \cV^{(n)}$ denoting the $n$-th level of the $\z$-expansion of $\cV$.
Moreover, one more piece of notation used throughout  this section
is the following: given a superfield $U(z)$, we denote 
$U|=U(z)|_{\S}$ to be its projection to the chiral superspace.

We focus on the computation of $E$
using equation (\ref{E-master3})
which in our case becomes 
\bea
\z\cdot \pa \ln E =
E_\hal{}^{\hat{\mu}}\F_{{\hat{\mu}} {\hat \b} }{}^{\hat \a} \z^{\hat \b}  
- \d_{\hat \m}{}^{\hat \a}
\big(E_{\hat \a}{}^{\hat \m} - \d_{\hat \a}{}^{\hat \m}\big)~.
\label{E-master3-chiral}
\eea
One should bear in mind that the connection now includes both 
the Lorentz and SU(2) terms, see Appendix A.
To determine the right hand side of (\ref{E-master3-chiral}) one needs  
to know special components of the connection, 
the vielbein and its inverse as functions of $\z$.
These can be found 
by solving iteratively, order-by-order in powers of $\z$,
the  equations\footnote{Equation (\ref{str-curvature1}) 
has to be used instead of  (\ref{str-curvature2}) in order 
to determine the vielbein at first order in $\z$. 
This follows from the fact that
 $(L_\z-1)L_\z E^{(1)}{}_\hm{}^A=0$. } 
(\ref{str-connection1})--(\ref{str-curvature2}).

One can notice several important simplifications 
even before starting to solve eqs.  (\ref{str-connection1})--(\ref{str-curvature2}). 
{}First of all, equation (\ref{eq-chiral-1}) tells us that 
\be 
E_{\hat{\mu}}{}^\ha=E_{\hal}{}^\hm=0~.
\ee
Second, since the structure group does not mix up the one-forms
$E^{\hat a}$ and $E^{\hat \a}$, the following identities hold:
$\F_\ha{}^\hbe=\F_\hal{}^\hb=R_\ha{}^\hbe=R_\hal{}^\hb=0$.
These results imply that eqs. (\ref{str-connection1})--(\ref{str-curvature2}) 
allow one to evaluate 
$E_{\hat{\mu}}{}^\hal,\,E_{\hal}{}^{\hat{\mu}}$ and $\F_{\hat{\mu}}{}_\hal{}^\hbe$
without knowing 
the other components of $E_M{}^A,\,E_A{}^M$ and $\F_M{}_A{}^B$.

Let us turn to evaluating $E_{\hat{\mu}}{}^\hal$ and 
$\F_{\hat{\mu}}{}_\hal{}^\hbe$ using
eqs. (\ref{str-connection1})--(\ref{str-curvature2}). 
According to the definition of the normal coordinate system, we have 
\bea
E_{\hat{\mu}}{}^\hal|=\d_{\hat{\mu}}{}^\hal~, \qquad
E_{\hal}{}^{\hat{\mu}}|=\d_{\hal}{}^{\hat{\mu}} ~, \qquad
\F_{\hat{\mu}}{}_\hal{}^\hbe|=0~.
\eea
Since $T_{\hat{\g}\hbe}{}^\hal =0$, 
equation (\ref{str-curvature1}) implies that 
\bea
E^{(1)}{}_{\hat{\mu}}{}^\hal=0~.
\label{viel-1}
\eea
Next, equation (\ref{str-connection1}) has the  following consequence:
\bea
(\z\cdot\pa+1)\F_{\hat{\mu}}{}_\hal{}^\hbe&=&
E_{\hat{\mu}}{}^\hde\z^\hga R_\hga{}_\hde{}_\hal{}^\hbe
\label{eq-connect}~.
\eea
To first order in $\z$, the latter gives
\bea
\F^{(1)}{}_{\hat{\mu}}{}_\hal{}^\hbe&=&
\hf\d_{\hat{\mu}}{}^\hde R_\hde{}_\hga{}_\hal{}^\hbe|\z^\hga~.
\eea
To compute $E_{\hat{\mu}}{}^\hal$ to second order in $\z$,  
it is handy to use equation (\ref{str-curvature2}) which gives
\bea
E^{(2)}{}_{\hat{\mu}}{}^\hal &=&
{1\over 6} \d_{\hat{\mu}}{}^\hde R_{\hde}{}_\hga{}_\hbe{}^\hal|\z^\hbe\z^\hga
\label{viel-2}~.
\eea
Next, making use of (\ref{viel-1}) and 
(\ref{eq-connect}) gives
\bea
\F^{(2)}{}_{\hat{\mu}}{}_\hal{}^\hbe &=&
\frac{1}{ 3}
\d_{\hat{\mu}}{}^\hde( \cD_{\hat{\rho}}R_\hde{}_\hga{}_\hal{}^\hbe)|\z^\hga\z^{\hat{\rho}}~.
\eea
Here we have used, for the first time,  the covariant Taylor expansion (\ref{Taylor2}) 
of the curvature. Further iterations lead to
\begin{subequations}
\bea
E^{(3)}{}_{\hat{\mu}}{}^\hal &=&
-{1\over 12} \d_{\hat{\mu}}{}^\hde(\cD_{\hat{\rho}} R_{\hde}{}_\hga{}_\hbe{}^\hal)|
\z^\hbe\z^\hga \z^{\hat{\rho}}~, \\
\F^{(3)}{}_{\hat{\mu}}{}_\hal{}^\hbe&=&
\frac{1}{8}\d_{\hat{\mu}}{}^\hta
\Big({1\over 3}  R_{\hta}{}_\hrh{}_\hde{}^{\hde'} R_{\hde'}{}_\hga{}_\hal{}^\hbe
+(\cD_{\hat{\rho}} \cD_{\hde}R_\hta{}_\hga{}_\hal{}^\hbe)\Big)\Big|
\z^\hga\z^{\hde}\z^{\hat{\rho}}~, \\
E^{(4)}{}_{\hat{\mu}}{}^\hal & = &
  {1\over 20}\d_{\hat{\mu}}{}^\hbe\Big(
 {1\over 6}R_{\hbe\hat{\tau}\hat{\rho}}{}^{\hde'}R_{\hde'\hde}{}_{\hga}{}^\hal
   + (\cD_{\hat{\tau}}\cD_{\hat{\rho}}R_{\hbe\hde}{}_{\hga}{}^\hal)
  \Big)\Big|
  \z^\hga\z^\hde \z^{\hat{\rho}}\z^{\hat{\tau}}~.
  \eea
\end{subequations}
As a result, we have computed  the components $E^{(n)}{}_{\hat{\mu}}{}^\hal$  of the vielbein, 
\bea
E_{\hat{\mu}}{}^{\hal}&=& \d_{\hat{\mu}}{}^\hal
+E^{(2)}{}_{\hat{\mu}}{}^\hal +E^{(3)}{}_{\hat{\mu}}{}^\hal
+E^{(4)}{}_{\hat{\mu}}{}^\hal~.
\label{viel}
\eea
Since  $E_{\hat{\mu}}{}^\ha=0$, 
the components $E_{\hal}{}^{\hat{\mu}}$ of the inverse  vielbein 
constitute  the inverse of the matrix (\ref{viel})
which can be easily computed.
Now, the master equation (\ref{E-master3-chiral}) becomes
\bea
\z\cdot\pa\ln E&=&
\d_\hal{}^{\hat{\mu}}\F^{(1)}{}_{\hat{\mu}}{}_\hbe{}^\hal{}\z^\hbe
+\d_\hal{}^{\hat{\mu}}\F^{(2)}{}_{\hat{\mu}}{}_\hbe{}^\hal{}\z^\hbe
+\d_\hal{}^{\hat{\mu}}\F^{(3)}{}_{\hat{\mu}}{}_\hbe{}^\hal{}\z^\hbe
-  \d_{\hal}{}^{\hat{\nu}} \d_{\hga}{}^{\hat{\mu}}E^{(2)}{}_{\hat{\nu}}{}^{\hga}
\F^{(1)}{}_{\hat{\mu}}{}_\hbe{}^\hal{}\z^\hbe
\non\\
&&
+\d_\hal{}^{\hat{\mu}}E^{(2)}{}_{\hat{\mu}}{}^{\hal}
-\d_\hal{}^{\hat{\mu}}\d_\hga{}^{\hat{\nu}}E^{(2)}{}_{\hat{\mu}}{}^{\hga} E^{(2)}{}_{\hat{\nu}}{}^{\hal}
+\d_\hal{}^{\hat{\mu}}E^{(3)}{}_{\hat{\mu}}{}^{\hal}
+\d_\hal{}^{\hat{\mu}}E^{(4)}{}_{\hat{\mu}}{}^{\hal}
~.~~~~~~
\label{compute-2}
\eea

At this stage, we need the explicit form of the curvature 
$R_\hal{}_\hbe{}_\hga{}^\hde$.
In accordance with (\ref{torsion-curvature}), it can be read off from 
the anticommutator 
$\{\cDB^\ad_i,\cDB^\bd_j\}$,  eq. (\ref{acr2}).  
\bea
R_\hal{}_\hbe{}_\hga{}^\hde&=&
R^\ad_i{\,}^\bd_j{\,}^\gd_k{\,}_\dd^l \non \\
&=&
\Big(4\bar{S}_{ij}\ve^{\gd(\ad}\d^{\bd)}_\dd\d_k^l
+2\ve_{ij}\ve^{\ad\bd}\bar{Y}^{\gd}{}_{\dd}\d_k^l
+2\ve_{ij}\ve^{\ad\bd}\bar{S}_k{}^{l}\d^\gd_\dd
+4\bar{Y}^{\ad\bd}\ve_{k(i}\d_{j)}^l\d^\gd_\dd
\Big)~, 
\label{curv-1}
\eea
and hence 
\bea
R_\hal{}_\hbe{}_\hga{}^\hal&=&
-4\bar{S}_{jk}\ve^{\bd\gd}
-4\bar{Y}^{\bd\gd}\ve_{jk}
~.
\label{curv-2}
\eea
Now, using (\ref{curv-2}), the relations
\bea
\z^\hal \z^\hbe
&=&
\hf\big(\ve^{ij}\z_{\ad\bd}-\ve_{\ad\bd}\z^{ij}\big)~,~~~~~~
\z_{\ad\bd}:=\z_{\ad k} \z_\bd^k=\z_{\bd\ad}~,~~~
\z^{ij}:=\z_\gd^i \z^{\gd j}=\z^{ji}~,
\\
&&~~~~~~~~~~~~
\z^\hal\z^\hbe\z^\hga=
{1\over 3}\ve^{jk}\ve_{\ad(\bd}\z_{\gd)q}\z^{iq}
-{1\over 3}\ve_{\bd\gd}\ve^{i(j}\z_{\ad q}\z^{k)q}
~,~~~~~~~~~~~~~~~
\eea
and the Bianchi identities (\ref{Bianchi-3/2-2}), 
one can prove that 
\bea
\d_\hal{}^{\hat{\mu}}E^{(3)}{}_{\hat{\mu}}{}^{\hal}=0~.
\eea
Then eq. (\ref{compute-2}) drastically simplifies 
\bea
\z\cdot\pa\ln E&=&
-{1\over 3} R_\hal{}_\hga{}_\hbe{}^\hal|\z^\hbe \z^\hga
+{1\over 45}R_\hal{}_{\hat{\tau}}{}_{\hat{\rho}}{}^\hde
R_\hde{}_\hga{}_\hbe{}^\hal|\z^\hbe\z^\hga\z^{\hat{\rho}}\z^{\hat{\tau}}
~.~~~~~~
\label{eq-3-3}
\eea
Making use of the relations (\ref{curv-1}) and  (\ref{curv-2}) along with  the identities
\begin{subequations}
\bea
&&\z^4:={1\over 3}\z^{ij}\z_{ij}~,~~~
\z^{ij}\z^{kl}=-\ve^{i(k}\ve^{l)j}\z^4~,~~~
\z_{\ad\bd}\z_{\gd\dd}=\ve_{\ad(\gd}\ve_{\dd)\bd}\z^4
~,~~~
\z_{\ad\bd}\z^{ij}=0~,~~~~~~
\\
&&~~~~~~~~~~~~~~~~~~
\z^\hal\z^\hbe\z^{\hga}\z^{\hde}
=
{1\over 4}\big(\ve^{ij}\ve^{kl}\ve_{\ad(\gd}\ve_{\dd)\bd}
-\ve_{\ad\bd}\ve_{\gd\dd}\ve^{i(k}\ve^{l)j}\big)\z^4~,
\eea
\end{subequations}
equation (\ref{eq-3-3}) becomes
\bea
\z\cdot\pa\ln E&=&
{4\over 3}\bar{Y}^{\ad\bd}|\z_{\ad\bd}
-{4\over 3}\bar{S}_{ij}|\z^{ij}
+{8\over 27}\big(
\bar{Y}^{\ad\bd}\bar{Y}_{\ad\bd}
-\bar{S}_{ij}\bar{S}^{ij}
\big)\big|
\z^4
~.~~~~~~
\eea
Its solution is given by  the simple formula
\bea
E&=&
\cE\Big(
1
+{2\over 3}\bar{Y}^{\ad\bd}|\z_{\ad\bd}
-{2\over 3}\bar{S}_{ij}|\z^{ij}
\Big)~,~~~~~~~~
\label{E-cE}
\eea
where $\cE={\rm Ber} \,(\cE_{\hat m}{}^{\hat a})$ is the chiral density.

Relation (\ref{E-cE}) can be used to reduce an integral over the full superspace 
to that over the chiral subspace.
Consider the functional
\bea
\int \rd^4 x \,{\rm d}^4\q{\rm d}^4{\bar \q}\,E\, U
=\int \rd^4 x \,{\rm d}^4\q {\rm d}^4\z\,E(\tilde{z}, \z) \, U(\tilde{z}, \z) ~,
\eea
where $U(z)$ is a scalar and isoscalar superfield, 
and $\tilde{z}^{\hat m}= (x^m, \q^\m_i$) the variables parametrizing the chiral subspace. 
In the normal coordinates, one represents 
$U$  by its covariant Taylor expansion in $\z$, 
eq. (\ref{Taylor2}), then evaluates the product $E\,U$,  
and finally performs the integration over $\rd^4\z$.
The result is as follows:
\bea
\int \rd^4 x \,{\rm d}^4\q{\rm d}^4{\bar \q}\,E\, U
= \int {\rm d}^4x \,{\rm d}^4 \q \, \cE \, \bar{\D} U \big|~.
\label{chiralproj1}
\eea
Here  $\bar{\D}$ denotes the following fourth-order operator: 
\bea
\bar{\D}
&=&\frac{1}{96} \Big((\cDB^{ij}+16\bar{S}^{ij})\cDB_{ij}
-(\cDB^{\ad\bd}-16\bar{Y}^{\ad\bd})\cDB_{\ad\bd} \Big)
\non\\
&=&\frac{1}{96} \Big(\cDB_{ij}(\cDB^{ij}+16\bar{S}^{ij})
-\cDB_{\ad\bd}(\cDB^{\ad\bd}-16\bar{Y}^{\ad\bd}) \Big)~,
\label{chiral-pr}
\eea
where we have defined
\bea
\cDB^{\ad\bd}:=\cDB^{(\ad}_k\cDB^{\bd)k}~,\qquad
\cDB_{ij}:=\cDB_{\gd(i}\cDB_{j)}^\gd~.
\eea
The operator $\bar{\D}$ is the $\cN=2$ covariantly chiral projector \cite{Muller}. 
Its fundamental property  is that $\bar{\D} U$ is covariantly chiral,
for any scalar and isoscalar superfield $U(z)$,
\be
{\bar \cD}^{\ad}_i \bar{\D} U =0~.
\ee
In section 5, we obtain a different representation for the chiral projector.

%%%%%%%%%%%%%%%%%%%%%%%%%%%%%%%%%%%%%%%%%%%%%%%%%%

\section{Density  formula in $\cN=2$ supergravity}
\setcounter{equation}{0}

In this section, the supergravity action (\ref{InvarAc}) is reduced to components using the principle
of projective invariance. We start by elaborating some auxiliary tools.

%%%%%%%%%%%%%%%%%%%%%%%%%%%%%%%%%%%%%%%%%%%%%%%%%
%%%%%%%%%%%%%%%%%%%%%%%%%%%%%%%%%%%%%%%%%%%%%%%%%

\subsection{Relating the superspace and the space-time covariant derivatives}

${}$For any superfield $U(z)$ we define its projection $U|$ to be 
the lowest component in the expansion of $U(x,\q, \bar \q)$ with respect to 
$\q$s and $\bar \q$s,
\bea
U (z)|:=U(x,\q, \bar \q)|_{\q={\bar \q}=0}~.
\eea
One can similarly define the projection of the covariant derivatives: 
\bea
\cD_A|:=E_A{}^M(z)|\pa_M+\hf\O_A{}^{bc}(z)|M_{bc}+\F_A{}^{kl}(z)|J_{kl}~.
\eea
More generally, given a gauge covariant operator of the form $\cD_{A_1} \dots \cD_{A_n}$, 
its projection $\big(\cD_{A_1} \dots \cD_{A_n}\big)\big|$ is defined as 
\bea
\Big( \big(\cD_{A_1} \dots \cD_{A_n}\big)\big| U \Big)\Big|
:=  \big(\cD_{A_1} \dots \cD_{A_n}U\big)\big|~, 
\eea
with $U$ an arbitrary tensor superfield.
The reader should keep in mind that the projection operation defined above differs from that 
used in section 3.

${}$In the case of the vector covariant derivatives, $\cD_a$, 
their projection can be represented in the form:
\bea
\cD_a|=\CD_a 
+\Psi_a{}^\g_k(x)\cD_\g^k|+\bar{\Psi}_a{}_\gd^k(x)\cDB^\gd_k|
+\phi_a{}^{kl}(x)J_{kl}~,~~~~~~
\label{cD_a-proj}
\eea
with $\CD_a$  a space-time covariant derivative,
\bea
\CD_a=e_a+\o_a~,\qquad 
e_a=e_a{}^m(x)\pa_m~,\quad
\o_a=\hf\o_a{}^{bc}(x)M_{bc}~.
\eea
Here we have introduced several component gauge fields
defined as follows:
\begin{subequations}
\bea
E_a{}^m(z)|&=&e_a{}^m(x)
+\Psi_a{}^\g_k(x)E_\g^k{}^m(z)|
+\bar{\Psi}_a{}_\gd^k(x)E^\gd_k{}^m(z)|
\label{no-gauge-1}
~,\\
E_a{}^\mu_r(z)|&=&\Psi_a{}^\g_k(x)E_\g^k{}^\mu_r(z)|
+\bar{\Psi}_a{}_\gd^k(x)E^\gd_k{}^\mu_r(z)|~,
\label{no-gauge-2}
\\
E_a{}_{\dot{\mu}}^r(z)|&=&\Psi_a{}^\g_k(x)E_\g^k{}_{\dot{\mu}}^r(z)|
+\bar{\Psi}_a{}_\gd^k(x)E^\gd_k{}_{\dot{\mu}}^r(z)|~,
\label{no-gauge-3}
\\
\O_a{}^{bc}(z)|&=&\o_{a}{}^{bc}(x)
+\Psi_a{}^\g_k(x)\O^k_\g{}^{bc}(z)|
+\bar{\Psi}_a{}_\gd^k\O_k^\gd{}^{bc}(z)|
~,\label{no-gauge-4}
\\
\F_a{}^{kl}(z)|&=&\phi_{a}{}^{kl}(x)
+\Psi_a{}^\b_j(x)\O^j_\b{}^{kl}(z)|
+\bar{\Psi}_a{}_\bd^j\O_j^\bd{}^{kl}(z)|
~.\label{no-gauge-5}
\eea
\end{subequations}
These include the inverse vielbein $e_a{}^m$, the Lorentz connection $\o_{a}{}^{bc}$
and the SU(2)-connection $\phi_{a}{}^{kl}$, as well as the gravitino fields
$\Psi_a{}^\g_k$ and $\bar{\Psi}_a{}_\gd^k$.

It is worth noting that if one chooses an $\cN=2$ analogue of  Wess-Zumino gauge \cite{WZ2} 
defined as
\bea
\cD_\a^i|={\pa\over\pa\q^\a_i}~,\qquad 
\cDB^\ad_i|={\pa\over\pa\qb_\ad^i}
~,
\eea
then the relations 
(\ref{no-gauge-1})--(\ref{no-gauge-5}) considerably simplify and take the form:
\begin{subequations}
\bea
&E_a{}^m(z)|=e_a{}^m(x)~,\qquad
E_a{}^\g_k(z)|=\Psi_a{}^\g_k(x)~,\qquad
E_a{}_\gd^k(z)|=\bar{\Psi}_a{}_\gd^k(x)~,\\
&
\O_a{}^{bc}(z)|=\o_{a}{}^{bc}(x)~,\qquad
\F_a{}^{kl}(z)|=\phi_{a}{}^{kl}(x)
~.
\eea
\end{subequations}

The space-time covariant derivatives obey the commutation relations
\bea
[\CD_a,\CD_b]&=&\cT_{ab}{}^c(x)\CD_c
+\hf\cR_{ab}{}^{cd}(x)M_{cd}~.
\eea
Here the torsion tensor determines the rule for integration by parts:
\bea
\int\rd^4x\, e\,\CD_a v^a=
\int\rd^4x\, e\,v^a \cT_a{}_b{}^b~, \qquad e^{-1}:= \det (e_a{}^m)~,
\label{integ-by-parts}
\eea
with $v^a$ an arbitrary vector field.

The space-time torsion $\cT_{ab}{}^c$ and curvature $\cR_{ab}{}^{cd}$
can be related to those appearing in the superspace (anti-)commutation relations
(\ref{acr1}--\ref{acr5}).  Using the definition (\ref{cD_a-proj}) and eqs. 
(\ref{acr1}--\ref{acr5}), one can 
evaluate the projection of 
the commutator $[\cD_a,\cD_b]$ to be 
\bea
&&[\cD_a,\cD_b]|=
\cT_{ab}{}^c\CD_c
-4\ri\Psi_{[a}{}^\g_k\bar{\Psi}_{b]}{}_\dd^k\CD_\g{}^\dd 
+\hf\cR_{ab}{}^{cd}M_{cd}
-\Psi_{[a}{}^\g_kR_{b]}{}_\g^k{}^{cd}|M_{cd}
-\bar{\Psi}_{[a}{}_\gd^kR_{b]}{}^\gd_k{}^{cd}|M_{cd}
\non
\\
&&~~~
+\hf\Psi_{[a}{}^\g_k\Psi_{b]}{}^\d_lR_\g^k{}_\d^l{}^{cd}|M_{cd}
+\hf\bar{\Psi}_{[a}{}_\gd^k\bar{\Psi}_{b]}{}_\dd^lR^\gd_k{}^\dd_l{}^{cd}|M_{cd}
+\Psi_{[a}{}^\g_k\bar{\Psi}_{b]}{}_\dd^lR_\g^k{}^\dd_l{}^{cd}|M_{cd}
+2(\CD_{[a}\Psi_{b]}{}^\g_k)\cD_\g^k|
\non
\\
&&~~~
-2\Psi_{[a}{}^\a_iT_{b]}{}_\a^i{}^\g_k\cD_\g^k|
-2\bar{\Psi}_{[a}{}_\ad^iT_{b]}{}^\ad_i{}^\g_k\cD_\g^k|
-2\phi_{[a}{}_k{}^{l}\Psi_{b]}{}^\g_l\cD_\g^k|
-4\ri\Psi_{[a}{}^\d_l\bar{\Psi}_{b]}{}_\dd^l\Psi_\d{}^\dd{}^\g_k\cD_\g^k|
\non
\\
&&~~~
+2(\CD_{[a}\bar{\Psi}_{b]}{}_\gd^k)\cDB^\gd_k|
-2\Psi_{[a}{}^\a_iT_{b]}{}_\a^i{}_\gd^k\cDB^\gd_k|
-2\bar{\Psi}_{[a}{}_\gd^iT_{b]}{}^\g_i{}_\gd^k\cDB^\gd_k|
+2\phi_{[a}{}^{k}{}_{l}\bar{\Psi}_{b]}{}_{\gd}^l\cDB^\gd_k|
\non
\\
&&~~~
-4\ri\Psi_{[a}{}^\d_l\bar{\Psi}_{b]}{}_\dd^l\bar{\Psi}_\d{}^\dd{}_\gd^k\cDB^\gd_k|
+2(\CD_{[a}\phi_{b]}{}^{kl})J_{kl}
-2\Psi_{[a}{}^\g_jR_{b]}{}_\g^j{}^{kl}|J_{kl}
-2\bar{\Psi}_{[a}{}_\gd^jR_{b]}{}^\gd_j{}^{kl}|J_{kl}
\non
\\
&&~~~
+\Psi_{[a}{}^\g_i\Psi_{b]}{}^\d_jR_\g^i{}_\d^j{}^{kl}|J_{kl}
+\bar{\Psi}_{[a}{}_\gd^i\bar{\Psi}_{b]}{}_\dd^jR^\gd_i{}^\dd_j{}^{kl}|J_{kl}
+2\Psi_{[a}{}^\g_i\bar{\Psi}_{b]}{}_\dd^jR_\g^i{}^\dd_j{}^{kl}|J_{kl}
+2\phi_{[a}{}^{k}{}_j\phi_{b]}{}^{jl}J_{kl}
\non
\\
&&~~~
-4\ri\Psi_{[a}{}^\g_j\bar{\Psi}_{b]}{}_\dd^j\phi_\g{}^\dd{}^{kl}J_{kl}~.
\eea
On the other hand, the commutator $[\cD_a,\cD_b]$
can be evaluated using eqs. (\ref{acr1}--\ref{acr5}).
Comparing the similar structures on both sides gives a number 
of important relations including the following:
\begin{subequations}
\bea
\cT_{ab}{}^c&=&
4\ri\Psi_{[a}{}^\g_k\bar{\Psi}_{b]}{}_\dd^k(\s^c)_\g{}^\dd~,
\label{T-1}
\\
(\CD_{[a}\Psi_{b]}{}^\g_k)&=&
\hf T_{ab}{}^\g_k|
+\Psi_{[a}{}^\a_iT_{b]}{}_\a^i{}^\g_k|
+\bar{\Psi}_{[a}{}_\ad^iT_{b]}{}^\ad_i{}^\g_k|
+\phi_{[a}{}_k{}^{l}\Psi_{b]}{}^\g_l
+2\ri\Psi_{[a}{}^\d_l\bar{\Psi}_{b]}{}_\dd^l\Psi _\d{}^\dd{}^\g_k
\label{D-Psi}
~,\\
(\CD_{[a}\bar{\Psi}_{b]}{}_\gd^k)
&=&
\hf T_{ab}{}_\gd^k|
+\Psi_{[a}{}^\a_iT_{b]}{}_\a^i{}_\gd^k|
+\bar{\Psi}_{[a}{}_\ad^iT_{b]}{}^\ad_i{}_\gd^k|
-\phi_{[a}{}^{k}{}_{l}\bar{\Psi}_{b]}{}_{\gd}^l
+2\ri\Psi_{[a}{}^\d_l\bar{\Psi}_{b]}{}_\dd^l\bar{\Psi}_\d{}^\dd{}_\gd^k~,
\label{D-Psi-bar}
\\
\cR_{ab}{}^{cd}&=&
R_{ab}{}^{cd}|
+2\Psi_{[a}{}^\g_kR_{b]}{}_\g^k{}^{cd}|
+2\bar{\Psi}_{[a}{}_\gd^kR_{b]}{}^\gd_k{}^{cd}|
-\Psi_{[a}{}^\g_k\Psi_{b]}{}^\d_lR_\g^k{}_\d^l{}^{cd}|
\non\\
&&
-\bar{\Psi}_{[a}{}_\gd^k\bar{\Psi}_{b]}{}_\dd^lR^\gd_k{}^\dd_l{}^{cd}|
-2\Psi_{[a}{}^\g_k\bar{\Psi}_{b]}{}_\dd^lR_\g^k{}^\dd_l{}^{cd}|
~,\\
(\CD_{[a}\phi_{b]}{}^{kl})
&=&
\hf R_{ab}{}^{kl}|
+\Psi_{[a}{}^\g_jR_{b]}{}_\g^j{}^{kl}|
+\bar{\Psi}_{[a}{}_\gd^jR_{b]}{}^\gd_j{}^{kl}|
-\hf\Psi_{[a}{}^\g_i\Psi_{b]}{}^\d_jR_\g^i{}_\d^j{}^{kl}|
-\hf\bar{\Psi}_{[a}{}_\gd^i\bar{\Psi}_{b]}{}_\dd^jR^\gd_i{}^\dd_j{}^{kl}|
\non\\
&&
-\Psi_{[a}{}^\g_i\bar{\Psi}_{b]}{}_\dd^jR_\g^i{}^\dd_j{}^{kl}|
-\phi_{[a}{}^{k}{}_j\phi_{b]}{}^{jl}
+2\ri\Psi_{[a}{}^\g_j\bar{\Psi}_{b]}{}_\dd^j\phi_\g{}^\dd{}^{kl}~.
\eea
\end{subequations}
In what  follows, we will often use eq. (\ref{T-1}), (\ref{D-Psi}) and (\ref{D-Psi-bar}).

\subsection{The component action }

We turn to demonstrating that the component reduction of action (\ref{InvarAc})  is 
\bea
S&=&
\oint_C {\rm d}  \mu^{(-2,-4)}
\int\rd^4 x \,e
\Big{[}
\frac{1}{ 16}({\cD}^-)^2(\cDB^-)^2
+\frac{3}{ 4}S^{--}(\cDB^-)^2
+\frac{3}{ 4}\bar{S}^{--}(\cD^{-})^2
+9S^{--}\bar{S}^{--}
\non\\
&&
+\frac{\ri }{4}\Psi^{\a\ad}{}_\a^{-}(\cD^-)^2\cDB_\ad^{-}
+\frac{\ri}{4}\bar{\Psi}^{\a\ad}{}_\ad^{-}(\cDB^{-})^2\cD_\a^{ -}
-\phi^{\a\ad}{}^{--}\cD_{\a }^-\cDB_{\ad}^- 
\non \\
&&
+(\s^{ab})^{\a\b} \Psi_{a}{}_\a^{-} \Big( \Psi_{b}{}_\b^{-}(\cD^-)^2
+2 \bar{\Psi}_{b}{}^{\bd-}\cD_\b^{-}\cDB_{\bd}^{-} \Big)
+(\ts^{ab})^{\ad\bd} 
\bar{\Psi}_a{}_{\ad}^{-} \Big(
\bar{\Psi}_b{}_\bd^{-} 
(\cDB^{-})^2
+2 \Psi_{b}{}^{\b-} \cD_\b^{-}\cDB_{\bd}^{-} \Big)
\non\\
&&
+3\ri \Big( \bar{\Psi}^{\a\ad}{}_\ad^{-}\bar{S}^{--}\cD_\a^{-}
+\Psi^{\a\ad}{}_\a^{-}S^{--}\cDB_\ad^{-}\Big)
-4 
\phi_{a}{}^{--} \Big(
(\s^{ab})^{\b\g}
\Psi_{b}{}_{\b}^-\cD_{\g}^-
-(\ts^{ab})^{\bd\gd}
\bar{\Psi}_{b}{}_{\bd}^-\cDB_\gd^{-}\Big) \non \\
&&
+4\ve^{abcd}(\s_d)_{\a\bd}\Psi_{a}{}^{\a-}\bar{\Psi}_{b}{}^{\bd-} \Big(
\Psi_{c}{}^{\g-}
\cD_\g^{-}
+\bar{\Psi}_c{}^{\gd-}\cDB_\gd^{-} \Big) 
-12\ve^{abcd}(\s_d)_{\a\bd}\Psi_{a}{}^{\a -}\bar{\Psi}_b{}^{\bd-}\phi_{c}{}^{--}
\non \\
&&+12(\s^{ab})^{\a\b}\Psi_a{}_\a^{-}\Psi_{b}{}_\b^{-}S^{--}
+12(\ts^{ab})^{\ad\bd}\bar{\Psi}_a{}_\ad^{-}\bar{\Psi}_b{}_{\bd}^{-}\bar{S}^{--}
\Big{]}\cL^{++}(z,u^+)\Big|~,
~~~~~~~~~
\label{Sfin-0}
\eea
where
 \be
 S^{\pm\pm}:=u^\pm_iu^\pm_j S^{ij}~,\qquad
\Psi_a{}_\a^{\pm}:=u^\pm_i\Psi_a{}_\a^i~, 
\qquad \phi_a{}^{\pm\pm}:=u^\pm_iu^\pm_j\phi_a{}^{ij}~, 
\ee
and similarly for  ${\bar S}^{\pm\pm}$ and $\bar{\Psi}_a{}_\ad^{\pm}$.
The spinor derivatives $\cD^-_\a$ and ${\bar \cD}^-_\ad$ 
are obtained from $\cD^+_\a$ and ${\bar \cD}^+_\ad$ 
defined in (\ref{ana-introduction}) by replacing 
$u^+_i \to u^-_i$.
The contour integration measure in (\ref{Sfin-0}) is defined as follows:
\be
{\rm d} \mu^{(-2,-4)}\equiv -{1\over 2\pi}{u_i^+\rd u^{+i}\over (u^+u^-)^4}
=  -{1\over 2\pi}{(\dt{u}^+ u^+) \over (u^+u^-)^4}\,{\rm d} t ~,
\label{measure}
\ee
with $t$ an evolution parameter along the contour $C$, 
and $\dt{f}:= \rd f(t)/ \rd t$ the time derivative of a function $f(t)$.
Here $u^-_i$ is a constant isotwistor subject only to the restriction 
that $u^-_i$ and $u^+_i(t)$ are linearly independent at each point 
of the closed contour $C$, that is $(u^+u^-) \neq 0$.
The remainder of this section is devoted to the derivation of (\ref{Sfin-0}).

In what follows, we often change  bases in the space of isotensors
by the rule $ A^i  \to A^\pm:=A^i u_i^\pm$ using the completeness relation
\bea
(u^+u^-)\,\d^i_j=
u^{+i}u^-_j-u^{-i}u^+_j~.
\eea
We also find it helpful to introduce  a notational convention that differs slightly from that 
used in \cite{KLRT-M,KT-Msugra1,KT-Msugra3}.
Specifically, $F^{(p,q)} (u^+,u^-)$ denotes a homogeneous function of $u^+$s and $u^-$s, 
with integers $p$ and $q$ being the corresponding degrees of homogeneity 
with  respect to $u^+$s and
$u^-$s, that is:  $F^{(p,q)} (c\,u^+,u^-)=c^p F^{(p,q)} (u^+,u^-)$ 
and $F^{(p,q)} (u^+,c\,u^-)=c^q F^{(p,q)} (u^+,u^-)$, where 
$c \in {\mathbb C} \setminus \{0\}$. 
This convention is reflected in the definition (\ref{measure}).
In the case of a homogeneous function of $u^+$s only, 
we use the simplified notation: $F^{(n)} (u^+) \equiv F^{(n,0)} (u^+)$; 
if $n>0$, we can also write $F^{(n)} \equiv F^{+\cdots +}$, 
where the number of $+$ superscripts is equal to $n$.
In the case of a homogeneous function of $u^-$s only, 
we often use the simplified notation $F^{-\dots -} (u^-) \equiv F^{(0,m)} (u^-)$
with $m>0$, where the number of $-$ superscripts is equal to $m$.

A few words are in order regarding our strategy of deriving (\ref{Sfin-0}).
It is clear that the component Lagrangian corresponding to the action (\ref{InvarAc}) 
should be a combination of terms with four and less spinor covariant derivatives 
acting on $\cL^{++}$. In the complete set 
of spinor covariant derivatives, 
$\cD^i_\a$ and ${\bar \cD}_{\ad }^i$, 
these derivatives should be linearly independent from 
the operators $\cD^+_\a$ and ${\bar \cD}^+_\ad$ which annihilate $\cL^{++}$.
A natural way to define such a subset of spinor covariant derivatives 
is to pick an isotwistor $u^-_i$ such that $(u^+u^-)\neq 0$.
Then the operators $\cD^-_\a$ and ${\bar \cD}^-_\ad$ clearly 
satisfy the required criterion. In other words, in order to construct the component 
action one is forced  to introduce an external isotwistor $u^-_i$ which does not show up 
in the original action  (\ref{InvarAc}).\footnote{This is similar to the Faddeev-Popov  
quantization of Yang-Mills theories. In order to develop a path-integral 
representation for the vacuum amplitue $\langle {out} | {in} \rangle$, 
one has to introduce a gauge fixing condition. 
However, the amplitue $\langle {out} | {in} \rangle$ must be independent of the 
gauge condition introduced.}
The latter involves only the isotwistor $u^+_i$, 
and is invariant under arbitrary re-scalings 
\be 
u_i^+(t)  \to c(t) \,u^+_i(t) ~, \qquad  c(t) \neq 0~,
\label{re-scale-u+}
\ee
along the  integration contour.
Therefore, the component action should be invariant under arbitrary 
projective transformations  (\ref{projectiveGaugeVar}). 
Indeed, the invariance 
under infinitesimal transformations of the form 
\be
u^-_i~\to~ u^-_i +\d u^-_i~, \qquad
\d u^-_i\,=\,\a(t)\,u^-_i+\b(t)\,u^+_i(t)~, 
\label{delta-u-}
\ee
implies independence of the action from the choice of $u^-_i$.
Since both $u^-_i$ and $\d {u}^-_i$ are required to  be time-independent, 
the transformation parameters  should obey the equations:
\bea
\dt{\a}=\b\,{(\dt{u}^+u^+)\over (u^+u^-)}~, \qquad \dt{\b}=-\b\,{(\dt{u}^+u^-)\over (u^+u^-)}~.
\label{ode}
\eea
Setting $\b=0$ in (\ref{delta-u-}) gives a scale transformation,  $\d u^-_i=\a\,u^-_i$.
Therefore, the component action must be invariant under arbitrary rigid re-scalings of $u^-_i$.
If the component Lagrangian density is chosen to be homogeneous in  
$u^-_i$ of degree zero, then  the invariance under rigid re-scalings of $u^-_i$
clearly extends to that under the time-dependent 
$\a$-transformations in (\ref{delta-u-}). It turns out that a nontrivial piece of information 
is provided by requiring the action to be invariant under the $\b$-transformations
in (\ref{delta-u-}).

On general grounds, it is not difficult to fix a four-derivative term 
in the component Lagrangian corresponding to the action (\ref{InvarAc}). 
We have 
\bea
S=S_0+\cdots~, \qquad 
S_0 = \frac{1}{ 16}\oint {\rm d} \mu^{(-2,-4)}
\int\rd^4 x \,e\,
({\cD}^-)^2(\cDB^-)^2
\cL^{++}(z,u^+)\Big|~,~~~~~~
\label{projectiveAnsatz0}
\eea
where the dots denote all the terms with at the most three spinor derivatives. 
The functional $S_0$ is obviously invariant under the 
local re-scalings of $u^+_i$, eq. (\ref{re-scale-u+}), 
and also under the $\a$-transformations in  (\ref{delta-u-}).
It turns out,  however, that $S_0$ is not invariant under the $\b$-transformation in  (\ref{delta-u-}).
To cancel out the $\b$-variation of $S_0$, it is necessary to add to $S_0$ 
some terms with three and less spinor derivatives acting on $\cL^{++}$. 
The latter produce new non-vanishing contributions of lower order under the
the $\b$-transformation in  (\ref{delta-u-}).
As a result, we end up with  a
well-defined  iterative procedure to restore a projective invariant action.
Conceptually, our approach below is quite simple.

Before proceeding with the computation, it is useful to collect some auxiliary results
and make a technical comment.
Since the superfield Lagrangian $\cL^{++}(z,u^+)$ 
is a  weight-two projective supermultiplet, it holds that 
\begin{subequations}
\bea
J_{kl}\cL^{++}&=&-{1\over (u^+u^-)}\Big(u^+_{(k}u^+_{l)}D^{(-1,1)}-2u^+_{(k}u^-_{l)}\Big)\cL^{++}~,
\label{JL-1}\\
{\rd\over \rd t}\cL^{++}
&=&2{(\dt{u}^+u^-)\over (u^+u^-)}\cL^{++}
-{(\dt{u}^+u^+)\over (u^+u^-)}D^{(-1,1)}\cL^{++}~,
\\
(\dt{u}^+u^+)J_{kl}\cL^{++}&=&
u^+_{(k}u^+_{l)}{\rd\over \rd t}\cL^{++}-2{(\dt{u}^+u^-)\over (u^+u^-)}u^+_{(k}u^+_{l)}\cL^{++}
+2{(\dt{u}^+u^+)\over (u^+u^-)}u^+_{(k}u^-_{l)}\cL^{++}~,~~~~~
\label{JL-3}
\eea
\end{subequations}
with $J_{kl}$ the SU(2) generators.
Here the operator $D^{(-1,1)}$ is defined in (\ref{5}).
Consider now 
any operator $\cO^{--}$, which is independent of $u^+$, 
${\pa\cO^{--}/\pa u^{+i}}=0$,
and is homogeneous in the variables $u^-_i$ of degree $+2$.
Using equations (\ref{JL-1}--\ref{JL-3}), one gets
\bea
{({\dot u^+}u^+)\over (u^+u^-)^4}\cO^{--}J^{--}\cL^{++}=
{\rd\over \rd t}\Big[{\cO^{--}\over (u^+u^-)^2}\cL^{++}\Big]~.
\eea
This implies the following relation:
\bea
\oint {\rm d}  \mu^{(-2,-4)}\,\cO^{--}J^{--}\cL^{++}&=&0~.
\eea
Due to the identities
\bea
{[}J_{kl},\cD^{\pm}_\a{]}={u^{\pm}_{(k}u^-_{l)}\over (u^+u^-)}\cD^+_\a
-{u^{\pm}_{(k}u^+_{l)}\over (u^+u^-)}\cD^-_\a~,&&~~~
{[}J_{kl},\cDB^{\pm}_\ad{]}={u^{\pm}_{(k}u^-_{l)}\over (u^+u^-)}\cDB^+_\ad
-{u^{\pm}_{(k}u^+_{l)}\over (u^+u^-)}\cDB^-_\ad~,~~~~~
\\
\{\cD_\a^-,\cDB_\ad^-\}=8G_{\a\ad}J^{--}~,&&~~~~~~
[J^{--},\cD_\a^-]=[J^{--},\cDB_\ad^-]=0~,
\eea
we also obtain
\bea
\oint {\rm d}  \mu^{(-2,-4)}\,(\cD^-)^2(\cDB^-)^2\cL^{++}&=&
\oint {\rm d}  \mu^{(-2,-4)}\,\cD^{\a-}(\cDB^-)^2\cD_\a^-\cL^{++}\non \\
=\oint {\rm d}  \mu^{(-2,-4)}\,(\cDB^-)^2(\cD^-)^2\cL^{++}
&=&
\oint {\rm d}  \mu^{(-2,-4)}\,\cDB_\ad^-(\cD^-)^2\cDB^{\ad-}\cL^{++}
~.~~~~~~
\eea
These identities justify the fact that $S_0$ is unambiguously defined.

Using equations (\ref{ode}) and  (\ref{JL-1}--\ref{JL-3}),  one can also prove 
(compare with the similar observation in  the 5D case  \cite{KT-Msugra1}) 
 the following result: for any operator $\cO^{(1,3)\,kl}$,  
which is an homogenous function of degrees $1$ and $3$ in $u^+_i$
and $u^-_i$, respectively,  it holds that
\bea
&& \oint {\rm d}  \mu^{(-2,-4)}\,\b\, \cO^{(1,3)kl}J_{kl}\cL^{++} \non \\
&=&
\oint  {\rm d}  \mu^{(-2,-4)}
\frac{\b}{ (u^+u^-)}\Bigg\{
4\cO^{(1,3)+-}\cL^{++}
+u^+_{k}u^+_{l}\Big(D^{(-1,1)}\cO^{(1,3)kl}\Big)\cL^{++}
\Bigg\}~.~~~~~~~~~
\label{usf-2}
\eea
This identity will often be used in what follows.

Let us consider the variation of $S_0$, eq. (\ref{projectiveAnsatz0}), 
under the infinitesimal projective transformation  (\ref{delta-u-}).
Since $\cD^+_\a\cL^{++}=\cDB^{+}_\ad\cL^{++}=0$, we obtain
\bea
\d S_0
&=&
{1\over 16}\oint {\rm d}  \mu^{(-2,-4)}\b
\int\rd^4 x \,e
\Big{[}
\{\cD^{\a+},\cD_\a^-\cDB_\ad^-\cDB^{\ad-}\}
+\cD^{\a-}[\cD_\a^+,\cDB_\ad^-\cDB^{\ad-}]
\non\\
&&
+\cD^{\a-}\cD_\a^-\{\cDB_\ad^+,\cDB^{\ad-}\}
\Big{]}\cL^{++}\Big|~,
\eea
which is equivalent to
\bea
\d S_0
&=&
\frac{1}{ 16}\oint {\rm d}  \mu^{(-2,-4)}\b
\int\rd^4 x \,e
\Big{[}
\{\cD^{\a+},\cD_\a^-\}\cDB_\ad^-\cDB^{\ad-}
+4\{\cD_\a^{+},\cDB_\ad^-\}\cD^{\a -}\cDB^{\ad-}
\non\\
&&
-4[\{\cD_\a^{-},\cDB_\ad^-\},\cD^{\a+}]\cDB^{\ad-}
-4[\{\cD^{\a +},\cD_\a^{-}\},\cDB_\ad^-]\cDB^{\ad-}
-2\cD^{\a -}[\{\cD_\a^{+},\cDB_\ad^-\},\cDB^{\ad-}]
\non\\
&&
+\cD^{\a-}\cD_\a^-\{\cDB_\ad^+,\cDB^{\ad-}\}
\Big{]}\cL^{++}\Big|~.
\eea
Here the (anti)commutators 
can be evaluated by 
making use of the algebra (\ref{acr1})--(\ref{acr5}).
As a next step, we systematically move the Lorentz and  SU(2) generators to the right 
and then use  the identity $M_{ab}\cL^{++}=0$
and eq. (\ref{usf-2}).
If in this process some spinor covariant derivatives 
$\cD_\a^+$ or $\cDB_\ad^+$ are produced, we push them
to the right until they hit $\cL^{++}$, and the latter contribution vanishes due to 
$\cD_\a^+\cL^{++}=\cDB_\ad^+\cL^{++}=0$.
We then find
\bea
\d S_0
&=&
\frac{1}{ 16}\oint {\rm d}  \mu^{(-2,-4)}\b
\int\rd^4 x \,e
\Big{[}
-8\ri\upm \cD_{\a\ad}\cD^{\a -}\cDB^{\ad-}
-24S^{+-}\cDB_\ad^-\cDB^{\ad-}
-24\bar{S}^{+-}\cD^{\a-}\cD_\a^-
\non\\
&&
-16\upm (\cDB_\ad^-\bar{W}^{\ad\dd})\cDB_\dd^{-}
-48(\cDB_\ad^-S^{+-})\cDB^{\ad-}
-56(\cD^{\b -}\bar{S}^{+-})\cD_\b^-
\non\\
&&
+8\upm (\cD^{\b -}{W}_{\b\g})\cD^{\g -}
+16\upm (\cDB^{\ad-}G_{\a\ad})\cD^{\a -}
\non\\
&&
-192S^{--}\bar{S}^{+-}
-32(\cD^{\b -}\cD_{\b}^-\bar{S}^{+-})
-16\upm (\cD^{\a -}\cDB^{\ad-}G_{\a\ad})
\Big{]}\cL^{++} \Big|~.
\eea
This expression can be simplified if one notices that 
the Bianchi identities (\ref{Bianchi-3/2-1})--(\ref{Bianchi-3/2-4}) imply 
\begin{subequations}
\bea
\cD_\a^+S^{--}&=&-2\cD_\a^-S^{+-}~,~~~~~~
\cD_{\a l}S^{-l}={3\over \upm }\cD_\a^-S^{+-}~,
\\
\cDB^{\ad -}G_{\a\ad}&=&
{1\over 4\upm }\cD_{\a}^+\bar{S}^{--}
+\hf\cD^{\g -}W_{\a\g}~,
\\
\cD^{\a-}\cD^{\b-}W_{\a\b}&=&0~,~~~~~~
\cD^{\a-}\cD_\a^-\bar{S}^{+-}=
4S^{+-}\bar{S}^{--}
-4S^{--}\bar{S}^{+-}~,
\\
\cD^{\a -}\cDB^{\ad-}G_{\a\ad}&=&
-{2\over \upm }S^{+-}\bar{S}^{--}
+{2\over \upm }S^{--}\bar{S}^{+-}
~,
\eea
\end{subequations}
along with complex conjugate relations.
We then end up with the following variation:
\bea
\d S_0
&=&
\oint {\rm d}  \mu^{(-2,-4)}\b
\int\rd^4 x \,e
\Big{[}
-{\ri\over 2}(u^+u^-)\cD_{\a\ad}\cD^{\a -}\cDB^{\ad-}
-{3\over 2}S^{+-}\cDB_\ad^-\cDB^{\ad-}
-{3\over 2}\bar{S}^{+-}\cD^{\a-}\cD_\a^-
\non\\
&&
-3(\cDB_\ad^-S^{+-})\cDB^{\ad-}
-3(\cD^{\a -}\bar{S}^{+-})\cD_\a^-
-(u^+u^-)(\cDB_\ad^-\bar{W}^{\ad\dd})\cDB_\dd^{-}
+(u^+u^-)(\cD^{\a -}{W}_{\a\b})\cD^{\b -}
\non\\
&&
-6S^{--}\bar{S}^{+-}
-6S^{+-}\bar{S}^{--}
\Big{]}\cL^{++} \Big|~.
\label{dS_0}
\eea
To cancel out the terms with two derivatives,  we add to 
$S_0$ the following structure:
\bea
S_1
&=&
\oint {\rm d}  \mu^{(-2,-4)}
\int\rd^4 x \,e
\Big{[}
{3\over 4}S^{--}(\cDB^-)^2
+{3\over 4}\bar{S}^{--}(\cD^{-})^2
\Big{]}\cL^{++} \Big|
~.
\label{S_1}
\eea
Its variation is
\bea
\d S_1
&=&
\oint {\rm d}  \mu^{(-2,-4)}\b
\int\rd^4 x \,e
\Big{[}
{3\over 2}S^{+-}(\cDB^-)^2
+{3\over 2}\bar{S}^{+-}(\cD^{-})^2
\non\\
&&
-12S^{--}\bar{S}^{+-}
-12\bar{S}^{--}{S}^{+-}
\Big{]}\cL^{++}\Big|
~,
\label{dS_1}
~~~~~~~~~
\eea
and therefore the functional $S_0+S_1$ varies as
\bea
\d (S_0+S_1)
&=&
\oint {\rm d}  \mu^{(-2,-4)}\b
\int\rd^4 x \,e
\Big{[}
-{\ri\over 2}(u^+u^-)\cD_{\a\ad}\cD^{\a -}\cDB^{\ad-}
-3(\cDB_\ad^-S^{+-})\cDB^{\ad-}
\non\\
&&
-3(\cD^{\a -}\bar{S}^{+-})\cD_\a^-
-(u^+u^-)(\cDB_\ad^-\bar{W}^{\ad\dd})\cDB_\dd^{-}
+(u^+u^-)(\cD^{\a -}{W}_{\a\b})\cD^{\b -}
\non\\
&&
-18S^{--}\bar{S}^{+-}
-18S^{+-}\bar{S}^{--}
\Big{]}\cL^{++}\Big|
~.
\label{dS_01}
\eea
To cancel the variation in the last line,  we have to add to the action another term
\bea
S_2
&=&
\oint {\rm d}  \mu^{(-2,-4)}
\int\rd^4 x \,e
\Big{[}
9S^{--}\bar{S}^{--}
\Big{]}\cL^{++}\Big|~.
\label{S_2}
\eea
As a result, the functional $S_0+S_1+S_2$ varies as
\bea
&&\d (S_0+S_1+S_2)
=
\oint {\rm d}  \mu^{(-2,-4)}\b
\int\rd^4 x \,e
\Big{[}
-{\ri\over 2}(u^+u^-)\cD_{\a\ad}\cD^{\a -}\cDB^{\ad-}
-3(\cDB_\ad^-S^{+-})\cDB^{\ad-}
\non\\
&&~~~
-3(\cD^{\a -}\bar{S}^{+-})\cD_\a^-
-(u^+u^-)(\cDB_\ad^-\bar{W}^{\ad\dd})\cDB_\dd^{-}
+(u^+u^-)(\cD^{\a -}{W}_{\a\b})\cD^{\b -}
\Big{]}\cL^{++}\Big|~.~~~~~~~~~~
\label{dS_012}
\eea
In the first term of the variation obtained, we can make use of 
 (\ref{cD_a-proj}). This leads to
\bea
&&\oint {\rm d}  \mu^{(-2,-4)}\b
\int\rd^4 x \,e\,\Big{[}
-{\ri\over 2}(u^+u^-)\cD_{\a\ad}\cD^{\a -}\cDB^{\ad-}\Big{]} \cL^{++}\Big|
\non\\
&&=
\oint {\rm d}  \mu^{(-2,-4)}\b
\int\rd^4 x \,e
\Big{[}-{\ri\over 2}(u^+u^-)\Big(
\CD_{\a\ad}
-{1\over\upm}\Psi_{\a\ad}{}^{\g+}\cD_\g^-
+{1\over\upm}\Psi_{\a\ad}{}^{\g-}\cD_\g^+
\non\\
&&~~~
+{1\over\upm}\bar{\Psi}_{\a\ad}{}_\gd^+\cDB^{\gd -}
-{1\over\upm}\bar{\Psi}_{\a\ad}{}_\gd^-\cDB^{\gd+}
+\phi_{\a\ad}{}^{kl}J_{kl}
\Big)
\cD^{\a -}\cDB^{\ad-}
\Big{]}
\cL^{++}\Big|~.~~~~~~
\label{eq-112}
\eea
This variation can be simplified, in complete analogy with the above calculation, 
by systematically moving the Lorentz and  SU(2) generators 
as well as the derivatives $\cD^+,\,\cDB^+$  to the right until they hit $\cL^{++}$, 
at which stage we can use the identity
$M_{ab}\cL^{++}=0$, eq. (\ref{usf-2}) and the analyticity conditions
$\cD^+_\a\cL^{++}=\cDB^{+}_\ad\cL^{++}=0$.
We then find
\bea
&&\oint {\rm d}  \mu^{(-2,-4)}\b
\int\rd^4 x \,e
\Big{[}
-{\ri\over 2}(u^+u^-)\cD_{\a\ad}\cD^{\a -}\cDB^{\ad-}
\Big{]}\cL^{++}\Big|
\non\\
&&=
\oint {\rm d}  \mu^{(-2,-4)}\b
\int\rd^4 x \,e
\Big{[}
-{\ri\over 2}(u^+u^-)\CD_{\a\ad}\cD^{\a -}\cDB^{\ad-}
-{\ri\over 4}\Psi^{\a\ad}{}_\a^{+}(\cD^-)^2\cDB_\ad^{-}
-{\ri\over 4}\bar{\Psi}^{\a\ad}{}_\ad^{+}(\cDB^{-})^2\cD_\a^{ -}
\non\\
&&
+2\phi_{\a\ad}{}^{+-}\cD^{\a -}\cDB^{\ad-}
+\phi_{\a\ad}{}^{--}\{\cD^{\a +},\cDB^{\ad-}\}
+\upm \Psi^{\a\ad}{}^{\g -}\cD_{\g\ad}\cD_\a^{ -}
+\upm \bar{\Psi}^{\a\ad}{}^{\gd-}\cD_{\a\gd}\cDB_\ad^{-}
\non\\
&&
+3{\ri}\upm \Psi^{\a\ad}{}^{\g -}Y_{\a\g}\cDB_\ad^{-}
-4{\ri}\Psi^{\a\ad}{}_\a^{-}S^{+-}\cDB_\ad^{-}
+{4\ri}\upm \bar{\Psi}^{\a\ad}{}^{\gd-}G_{\a\gd}\cDB_\ad^{-}
\non\\
&&
-{2\ri}\upm \bar{\Psi}^{\a\ad}{}_{(\ad}^-G_{\a\gd)}\cDB^{\gd -}
+3{\ri}\upm \bar{\Psi}^{\a\ad}{}^{\gd-}\bar{Y}_{\ad\gd}\cD_\a^{ -}
-4\ri\bar{\Psi}^{\a\ad}{}_\ad^{-}\bar{S}^{+-}\cD_\a^{-}
\non\\
&&
-{4\ri}\upm \Psi^{\a\ad}{}^{\g -}G_{\g\ad}\cD_\a^{ -}
+2\ri\upm \Psi^{\a\ad}{}_{(\a}^{-}G_{\b)\ad}\cD^{\b -}
+3\ri\upm \Psi^{\a\ad}{}^{\g -}(\cDB_\ad^{-}Y_{\a\g})
\non\\
&&
+3\ri\upm \bar{\Psi}^{\a\ad}{}^{\gd-}(\cD_\a^{ -}\bar{Y}_{\ad\gd})
-\ri\upm \Psi^{\a\ad}{}_\a^{-}(\cDB^{\dd -}\bar{W}_{\ad\dd})
-\ri\upm \bar{\Psi}^{\a\ad}{}_\ad^{-}(\cD^{-\d}W_{\a\d})
\non\\
&&
-3\ri\Psi^{\a\ad}{}_\a^{-}(\cDB_{\ad}^{-}S^{+-})
-3\ri\bar{\Psi}^{\a\ad}{}_\ad^{-}(\cD_\a^{-}\bar{S}^{+-})
\Big{]}\cL^{++}\Big|~.
\label{D_a-D^al-DB^ad}
\eea
Now,  in order to cancel out the second, third, fourth and fifth terms, 
 we have to add to the action one more term
\bea
S_3&=&
\oint {\rm d}  \mu^{(-2,-4)}
\int\rd^4 x \,e
\Big{[}\,
{\ri\over 4}\Psi^{\a\ad}{}_\a^{-}(\cD^-)^2\cDB_\ad^{-}
+{\ri\over 4}\bar{\Psi}^{\a\ad}{}_\ad^{-}(\cDB^{-})^2\cD_\a^{ -}
\non\\
&&
-\phi_{\a\ad}{}^{--}\cD^{\a -}\cDB^{\ad-}
\Big{]}\cL^{++}\Big|
\label{S_3}~.
\eea
Evaluating the variation of $S_3$ and combining it with 
$\d(S_0+S_1+S_2)$  gives
\bea
&&\d(S_0+S_1+S_2+S_3)=
\oint {\rm d}  \mu^{(-2,-4)}\b
\int\rd^4 x \,e
\Big{[}
-{\ri\over 2}(u^+u^-)\CD_{\a\ad}\cD^{\a -}\cDB^{\ad-}
\allowdisplaybreaks
\non\\
&&
+\upm \Psi^{\a\ad}{}^{\g -}\cD_{\g\ad}\cD_\a^{ -}
+\upm \Psi^{\a\ad}{}_\a^{-}\cD_{\b\ad}\cD^{\b-}
+\upm \bar{\Psi}^{\a\ad}{}^{\gd-}\cD_{\a\gd}\cDB_\ad^{-}
\allowdisplaybreaks
\non\\
&&
+\upm \bar{\Psi}^{\a\ad}{}_\ad^{-}\cD_{\a\bd}\cDB^{\bd-}
+3{\ri}\upm \Psi^{\a\ad}{}^{\g -}Y_{\a\g}\cDB_\ad^{-}
-9{\ri}\Psi^{\a\ad}{}_\a^{-}S^{+-}\cDB_\ad^{-}
\allowdisplaybreaks
\non\\
&&
+{4\ri}\upm \bar{\Psi}^{\a\ad}{}^{\gd-}G_{\a\gd}\cDB_\ad^{-}
-{\ri}\upm \bar{\Psi}^{\a\ad}{}_{\gd}^-G_{\a\ad}\cDB^{\gd -}
-\ri\upm \Psi^{\a\ad}{}_\a^{-}\bar{W}_{\gd\ad}\cDB^{\gd -}
\allowdisplaybreaks
\non\\
&&
-3(\cDB_\ad^-S^{+-})\cDB^{\ad-}
-(u^+u^-)(\cDB_\ad^-\bar{W}^{\ad\dd})\cDB_\dd^{-}
+3{\ri}\upm \bar{\Psi}^{\a\ad}{}^{\gd-}\bar{Y}_{\ad\gd}\cD_\a^{ -}
-9\ri\bar{\Psi}^{\a\ad}{}_\ad^{-}\bar{S}^{+-}\cD_\a^{-}
\allowdisplaybreaks
\non\\
&&
-{4\ri}\upm \Psi^{\a\ad}{}^{\g -}G_{\g\ad}\cD_\a^{ -}
+\ri\upm \Psi^{\a\ad}{}_{\b}^{-}G_{\a\ad}\cD^{\b -}
-\ri\upm \bar{\Psi}^{\a\ad}{}_\ad^{-} {W}_{\a\g}\cD^{\g-}
\allowdisplaybreaks
\non\\
&&
-3(\cD^{\a -}\bar{S}^{+-})\cD_\a^-
+(u^+u^-)(\cD^{\a -}{W}_{\a\b})\cD^{\b -}
+3\ri\upm \Psi^{\a\ad}{}^{\g -}(\cDB_\ad^{-}Y_{\a\g})
\allowdisplaybreaks
\non\\
&&
+3\ri\upm \bar{\Psi}^{\a\ad}{}^{\gd-}(\cD_\a^{ -}\bar{Y}_{\ad\gd})
-3\ri\upm \Psi^{\a\ad}{}_\a^{-}(\cDB^{\bd -}\bar{W}_{\ad\bd})
-3\ri\upm \bar{\Psi}^{\a\ad}{}_\ad^{-}(\cD^{-\b}W_{\a\b})
\allowdisplaybreaks
\non\\
&&
-9\ri\Psi^{\a\ad}{}_\a^{-}(\cDB_{\ad}^{-}S^{+-})
-9\ri\bar{\Psi}^{\a\ad}{}_\ad^{-}(\cD_\a^{-}\bar{S}^{+-})
\Big{]}\cL^{++}\Big|
~.
\label{dS_0123}
\eea
Let us consider the first to fifth terms in (\ref{dS_0123}) which involve vector 
covariant derivatives.  In this sector, we  apply  (\ref{cD_a-proj}), 
the formula for integration by parts, eq. (\ref{integ-by-parts}),  
with the space-time torsion (\ref{T-1})  expressed as 
\bea
\cT_{ab}{}^c&=&
-{4\ri\over (u^+u^-)}\Big(\Psi_{[a}{}^{\g +}\bar{\Psi}_{b]}{}_{\dd}^-(\s^c)_\g{}^\dd
-\Psi_{[a}{}^{\g -}\bar{\Psi}_{b]}{}_{\dd}^+(\s^c)_\g{}^\dd\Big)~.~~~~~~
\eea
Implementing also the usual iterative procedure, we obtain
\bea \allowdisplaybreaks
&&
\oint {\rm d}  \mu^{(-2,-4)}\b
\int\rd^4 x \, e\,(u^+u^-)
\Big{[}
-{\ri\over 2}
\CD_{\a\ad}\cD^{\a -}\cDB^{\ad-}
+ \Psi^{\a\ad}{}^{\b -}\cD_{\b\ad}\cD_\a^{ -}
+\Psi^{\a\ad}{}_\a^{-}\cD_{\b\ad}\cD^{\b-}
\non 
\allowdisplaybreaks
\\
&&~~~
+ \bar{\Psi}^{\a\ad}{}^{\gd-}\cD_{\a\gd}\cDB_\ad^{-}
+\bar{\Psi}^{\a\ad}{}_\ad^{-}\cD_{\a\bd}\cDB^{\bd-}
\Big{]}\cL^{++}\Big|
\non
\allowdisplaybreaks
\\
&&
=\oint {\rm d}  \mu^{(-2,-4)}\b
\int\rd^4 x \,e
\Big{[}
2(\s^{ab})^{\a}{}_{\b}\Psi_{[a}{}^{\b+}\bar{\Psi}_{b]}{}^{\ad-}\cD_\a^{-}\cDB_{\ad}^{-}
+2(\ts^{ab})^{\ad}{}_{\bd}\Psi_{[a}{}^{\a+}\bar{\Psi}_{b]}{}^{\bd-}\cD_\a^{-}\cDB_{\ad}^{-}
\non\\
&&~~~
+2(\s^{ab})^{\a}{}_{\b}\Psi_{[a}{}^{\b-}\bar{\Psi}_{b]}{}^{\ad+}\cD_\a^{-}\cDB_{\ad}^{-}
+2(\ts^{ab})^{\ad}{}_{\bd}\Psi_{[a}{}^{\a-}\bar{\Psi}_{b]}{}^{\bd+}\cD_\a^{-}\cDB_{\ad}^{-}
-2(\s^{ab})^{\a\b}\Psi_a{}_\a^{-}\Psi_{b}{}_\b^{+}(\cD^-)^2
\non\\
&&~~~
-2(\ts^{ab})^{\ad\bd}\bar{\Psi}_a{}_\ad^{-}\bar{\Psi}_b{}_{\bd}^{+}(\cDB^{-})^2
-4\upm (\s^{ab})_\b{}^\g\cT_{[a}{}_{|c|}{}^c\Psi_{b]}{}^{\b -}\cD_\g^{ -}
\non
\allowdisplaybreaks
\\
&&~~~
+4\upm (\ts^{ab})_\bd{}^\gd\cT_{[a}{}_{|c|}{}^c\bar{\Psi}_{b]}{}^{\bd-}\cDB_\gd^{-}
+4\upm (\s^{ab})_\b{}^\g(\CD_{[a}\Psi_{b]}{}^{\b -})\cD_\g^{ -}
\non\\
&&~~~
-4\upm (\ts^{ab})_\bd{}^\gd(\CD_{[a}\bar{\Psi}_{b]}{}^{\bd-})\cDB_\gd^{-}
+12(\s^{ab})_\b{}^\g\Psi_a{}^{\b -}\phi_{b}{}^{+-}\cD_\g^{ -}
-12(\ts^{ab})_\bd{}^\gd\bar{\Psi}_a{}^{\bd-}\phi_b{}^{+-}\cDB_\gd^{-}
\non
\allowdisplaybreaks
\\
&&~~~
+4(\s^{ab})^\a{}_\b\Psi_a{}^{\b -}\bar{\Psi}_b{}^{\gd-}\{\cDB_\gd^{+},\cD_\a^{ -}\}
-4(\ts^{ab})^\ad{}_\bd\bar{\Psi}_a{}^{\bd-}\Psi_b{}^{\g-}\{\cD_\g^+,\cDB_\ad^{-}\}
\non
\allowdisplaybreaks
\\
&&~~~
+4(\s^{ab})^\a{}_\b\Psi_a{}^{\b -}\Psi_b{}^{\g-}\{\cD_\g^+,\cD_\a^{ -}\}
-4(\ts^{ab})^\ad{}_\bd\bar{\Psi}_a{}^{\bd-}\bar{\Psi}_b{}^{\gd-}\{\cDB_\gd^{+},\cDB_\ad^{-}\}
\non
\allowdisplaybreaks
\\
&&~~~
-32\upm (\s^{ab})^\a{}_\b\Psi_a{}^{\b -}\bar{\Psi}_b{}^{\ad-}G_{\a\ad}
-8\upm (\s^{ab})^\a{}_\b\Psi_a{}^{\b -}\Psi_b{}^{\g-}Y_{\a\g}
\non
\allowdisplaybreaks
\\
&&~~~
-8\upm (\ts^{ab})^\ad{}_\bd\bar{\Psi}_a{}^{\bd-}\bar{\Psi}_b{}^{\gd-}\bar{Y}_{\ad\gd}
\Big{]}\cL^{++}\Big|~.
\label{122}
\eea
In order to cancel the first six terms in  (\ref{122}), 
we have to add to the action one more structure
\bea
S_4&=&\oint {\rm d}  \mu^{(-2,-4)}
\int\rd^4 x \,e
\Big{[}
-2(\s^{ab})^{\a}{}_{\b}\Psi_{[a}{}^{\b-}\bar{\Psi}_{b]}{}^{\ad-}\cD_\a^{-}\cDB_{\ad}^{-}
-2(\ts^{ab})^{\ad}{}_{\bd}\Psi_{[a}{}^{\a-}\bar{\Psi}_{b]}{}^{\bd-}\cD_\a^{-}\cDB_{\ad}^{-}
\non\\
&&
+(\s^{ab})^{\a\b}\Psi_a{}_\a^{-}\Psi_{b}{}_\b^{-}(\cD^-)^2
+(\ts^{ab})^{\ad\bd}\bar{\Psi}_a{}_\ad^{-}\bar{\Psi}_b{}_{\bd}^{-}(\cDB^{-})^2
\Big{]}\cL^{++}(z,u^+)\Big|
\label{S_4}
\eea
and consider the variation
$\d(S_0+S_1+S_2+S_3+S_4)$.
We use (\ref{cD_a-proj}), then move $\cD^+,\,\cDB^+$ derivatives, 
Lorentz and SU(2) generators  to the right. Next we should move to the left 
all  $\CD_a$ derivatives and use the rule for integration by parts, eq. (\ref{integ-by-parts}).
At this stage, we can use the identities
\bea
&&\CD_{[a}\Psi_{b]}{}^{\g-}=
-{\frac18}(\ts_{ab})^{\ad\bd}(\cD^{\g-}\bar{Y}_{\ad\bd})|
+{\frac18}(\s_{ab})^{\a\b}(\cD^{\g-}W_{\a\b})|
+{\frac14}(\s_{ab})^{\g\d}\cD_{\d}^-\bar{S}^{+-}|
\non\\
&&~~~
+\ri(\s_{[a})^{(\a}{}_{\bd}\Psi_{b]}{}_\a^{-}G^{\g)\bd}|
-{\ri\over2(u^+u^-)}({\ts}_{[a})^{\ad\g}\bar{\Psi}_{b]}{}_\ad^+\bar{S}^{--}|
+{\ri\over 2(u^+u^-)}({\ts}_{[a})^{\ad\g}\bar{\Psi}_{b]}{}_\ad^-\bar{S}^{+-}|
\non\\
&&~~~
-{\ri\over 2}({\s}_{[a})_\a{}^{\ad}\bar{\Psi}_{b]}{}_\ad^-{W}^{\a\g}|
-{\ri\over 2}({\s}_{[a})^{\g}{}_\bd\bar{\Psi}_{b]}{}_\ad^-\bar{Y}^{\ad\bd}|
+\frac{1}{ (u^+u^-)} \Big\{ \phi_{[a}{}^{+-}\Psi_{b]}{}^{\g -}
- \phi_{[a}{}^{--}\Psi_{b]}{}^{\g +} \Big\}~~~~~
\non\\
&&~~~
-\frac{2\ri}{ (u^+u^-)}(\s^c)_\d{}^\dd\Psi_c{}^{\g-}
\Big\{ \Psi_{[a}{}^{\d +}\bar{\Psi}_{b]}{}_\dd^-
-\Psi_{[a}{}^{\d -}\bar{\Psi}_{b]}{}_\dd^+ \Big\}
\label{D-Psi-2-2}
~~~~~~~~~~
\eea
and
\bea
&&\CD_{[a}\bar{\Psi}_{b]}{}_\gd^-
=
-{\frac18}(\s_{ab})^{\a\b}(\cDB_\gd^{-}Y_{\a\b})|
+{\frac18}(\ts_{ab})^{\ad\bd}(\cDB_{\gd}^{-}\bar{W}_{\ad\bd})|
-{\frac14}(\ts_{ab})_{\gd\dd}(\cDB^{\dd-}S^{+-})|
\non\\
&&~~~
-\ri(\s_{[a})^\a{}_{(\ad}\bar{\Psi}_{b]}{}^{\ad-}G_{\a\gd)}|
+{\ri\over2(u^+u^-)}({\s}_{[a})_{\a\gd}\Psi_{b]}{}^{\a +}S^{--}|
-{\ri\over2(u^+u^-)}({\s}_{[a})_{\a\gd}\Psi_{b]}{}^{\a -}S^{+-}|
\non\\
&&~~~
-{\ri\over2}({\s}_{[a})_\a{}^{\dd}\Psi_{b]}{}^{\a -}\bar{W}_{\dd\gd}|
-{\ri\over2}({\s}_{[a})^{\b}{}_\gd \Psi_{b]}{}^{\a -}Y_{\a\b}|
+\frac{1}{ (u^+u^-)}\Big\{  \phi_{[a}{}^{+-}\bar{\Psi}_{b]}{}_{\gd}^-
- \phi_{[a}{}^{--}\bar{\Psi}_{b]}{}_{\gd}^+ \Big\}
\non\\
&&~~~
-\frac{2\ri }{ (u^+u^-)}(\s^c)_\d{}^\dd
\Big\{
\Psi_{[a}{}^{\d +}\bar{\Psi}_{b]}{}_\dd^-
-\Psi_{[a}{}^{\d -}\bar{\Psi}_{b]}{}_\dd^+ \Big\}
\bar{\Psi}_c{}_\gd^- ~,
\label{D-Psi-bar-2-2}
\eea
which follow from (\ref{D-Psi}) and (\ref{D-Psi-bar}).
After rather long computation, which involves 
algebraic manipulations using some results 
from  Appendix A,  non-trivial cancellations occur. 
One  obtains
\bea
&&\d(S_0+S_1+S_2+S_3+S_4)=
\oint {\rm d}  \mu^{(-2,-4)}\b
\int\rd^4 x \,e
\Big{[}
-24(\s^{ab})^{\a\b}\Psi_a{}_\a^{-}\Psi_{b}{}_\b^{-}S^{+-}
\non
\allowdisplaybreaks
\\
&&
-24(\s^{ab})_{\a\b}\Psi_{a}{}^{\a +}\Psi_b{}^{\b -}S^{--}
-24(\ts^{ab})_{\ad\bd}\bar{\Psi}_a{}^{\ad-}\bar{\Psi}_b{}^{\bd-}\bar{S}^{+-}
-24(\ts^{cd})_{\ad\bd}\bar{\Psi}_{a}{}^{\bd+}\bar{\Psi}_b{}^{\ad-}\bar{S}^{--}
\non
\allowdisplaybreaks
\\
&&
-6\ri\bar{\Psi}^{\a\ad}{}_\ad^{-}\bar{S}^{+-}\cD_\a^{-}
-3\ri\bar{\Psi}_{\a\ad}{}^{\ad+}\bar{S}^{--}\cD^{\a-}
-6{\ri}\Psi^{\a\ad}{}_\a^{-}S^{+-}\cDB_\ad^{-}
-3\ri\Psi^{\a\ad}{}_\a^{+}S^{--}\cDB_\ad^{-}
\non
\allowdisplaybreaks
\\
&&
+8(\s^{ab})_{\a\b}\phi_{a}{}^{+-}\Psi_b{}^{\a -}\cD^{\b-}
+4(\s^{ab})_{\a\b}\phi_{a}{}^{--}\Psi_{b}{}^{\a +}\cD^{\b-}
-8(\ts^{ab})_{\ad\bd}\phi_a{}^{+-}\bar{\Psi}_b{}^{\ad-}\cDB^{\bd-}
\non
\allowdisplaybreaks
\\
&&
-4(\ts^{ab})_{\ad\bd}\phi_{a}{}^{--}\bar{\Psi}_{b}{}^{\ad+}\cDB^{\bd-}
-4\ve^{abcm}(\s_m)_{\a\ad}\Psi_{a}{}^{\a +}\Psi_{b}{}^{\b-}\bar{\Psi}_{c}{}^{\ad-}\cD_\b^{-}
\non
\allowdisplaybreaks
\\
&&
-4\ve^{abcm}(\s_m)_{\a\ad}\Psi_{a}{}^{\b-}\Psi_{b}{}^{\a -}\bar{\Psi}_{c}{}^{\ad+}\cD_\b^{-}
-4\ve^{abcm}(\s_m)_{\a\ad}\Psi_a{}^{\a-}\Psi_b{}^{\b+}\bar{\Psi}_c{}^{\ad-}\cD_\b^-
\non
\allowdisplaybreaks
\\
&&
-4\ve^{abcm}(\s_m)_{\a\ad}\Psi_{a}{}^{\a+}\bar{\Psi}_{b}{}^{\ad-}\bar{\Psi}_{c}{}^{\bd-}\cDB_\bd^{-}
-4\ve^{abcm}(\s_m)_{\a\ad}\Psi_{a}{}^{\a-}\bar{\Psi}_{b}{}^{\ad+}\bar{\Psi}_{c}{}^{\bd-}\cDB_\bd^{-}
\non
\allowdisplaybreaks
\\
&&
-4\ve^{abcm}(\s_m)_{\a\ad}\Psi_a{}^{\a -}\bar{\Psi}_b{}^{\ad-}\bar{\Psi}_c{}^{\bd+}\cDB_\bd^{-}
+12\ve^{abcm}(\s_m)_{\a\ad}\phi_{a}{}^{--}\Psi_{b}{}^{\a +}\bar{\Psi}_c{}^{\ad-}
\non
\allowdisplaybreaks
\\
&&
+12\ve^{abcm}(\s_m)_{\a\ad}\phi_{a}{}^{--}\Psi_b{}^{\a -}\bar{\Psi}_{c}{}^{\ad+}
+24\ve^{abcm}(\s_m)_{\a\ad}\phi_a{}^{+-}\Psi_b{}^{\a -}\bar{\Psi}_c{}^{\ad-}
\Big{]}\cL^{++}\Big|~.
\label{dS_01234}
\eea
The nontrivial point is that all terms with four gravitinos 
have identically cancelled out
at this stage. And one more iteration -- 
we have to add to the action the following structure:
\bea
S_5&=&
\oint {\rm d}  \mu^{(-2,-4)}
\int\rd^4 x \,e
\Big{[}
3\ri\bar{\Psi}^{\a\ad}{}_\ad^{-}\bar{S}^{--}\cD_\a^{-}
+3{\ri}\Psi^{\a\ad}{}_\a^{-}S^{--}\cDB_\ad^{-}
+12(\s^{ab})^{\a\b}\Psi_a{}_\a^{-}\Psi_{b}{}_\b^{-}S^{--}
\non\\
&&
+12(\ts^{ab})^{\ad\bd}\bar{\Psi}_a{}_\ad^{-}\bar{\Psi}_b{}_{\bd}^{-}\bar{S}^{--}
-4(\s^{ab})_{\b\g}\phi_{a}{}^{--}\Psi_{b}{}^{\g-}\cD^{\b-}
+4(\ts^{ab})^{\bd\gd}\phi_{a}{}^{--}\bar{\Psi}_{b}{}_{\gd}^-\cDB_\bd^{-}
\non\\
&&
-12\ve^{abcd}(\s_d)_{\g\ad}\phi_{a}{}^{--}\Psi_{b}{}^{\g -}\bar{\Psi}_c{}^{\ad-}
+4\ve^{abcd}(\s_d)_{\a\ad}\Psi_{a}{}^{\a-}\Psi_{b}{}^{\b-}\bar{\Psi}_{c}{}^{\ad-}\cD_\b^{-}
\non\\
&&
+4\ve^{abcd}(\s_d)_{\a\ad}\Psi_a{}^{\a-}\bar{\Psi}_b{}^{\ad-}\bar{\Psi}_c{}^{\bd-}\cDB_\bd^{-}
\Big{]}\cL^{++}\Big|~.
\label{S_5}
~~~~~~~~~~~~
\eea
This proves to complete the procedure. 
One can now check that 
\bea
&&\d(S_0+S_1+S_2+S_3+S_4+S_5)=\d S=0
\label{dS_012345}
\eea
We have thus demonstrated that the action (\ref{Sfin-0}) is uniquely 
obtained from the requirement of projective invariance.

\subsection{Analysis of the results} 

The component action (\ref{Sfin-0}) is the main result of this work.
In technical terms, our procedure for deriving   (\ref{Sfin-0}) from the original superspace 
action (\ref{InvarAc}) has many similarities with 
the earlier construction for 5D $\cN=1$  supergravity \cite{KT-Msugra1}.
There is, however, a very important  conceptual difference. 
The point is that, unlike the five dimensional  analysis in \cite{KT-Msugra1}, 
no Wess-Zumino gauge has been assumed 
in the process of deriving (\ref{Sfin-0}).\footnote{A careful analysis of the 5D construction 
\cite{KT-Msugra1} shows that the choice of the Wess-Zumino gauge was not essential.
It is just an unfortunate stereotype forced upon us by textbook lessons \cite{WB,GGRS,BK}
that choosing Wess-Zumino gauge is imperative for component reduction.}
In other words, all the gauge symmetries of the 
parental superspace action (\ref{InvarAc}) are preserved by its component counterpart
 (\ref{Sfin-0}).\footnote{Most of purely  gauge degrees of freedom are contained 
 in the vielbein and connection superfields for $\cD^i_\a$ and ${\bar \cD}^i_\ad$. 
In the construction used, however, these objects do not show up explicitly.}
This huge gauge freedom can be used at will 
depending on concrete dynamical circumstances.  
It is worth giving two examples.

The action  is invariant under the super-Weyl transformations generated by a covariantly 
chiral parameter $\s$, ${\bar \cD}^i_\ad \s =0$.
This local symmetry can be used to choose a useful gauge condition, 
for instance, to set 
the field strength $W$ of the compensating vector multiplet to be
\be
W=1~.
\ee

The action is invariant under local SU(2) transformations generated 
by a real symmetric parameter $K^{ij}$ that is otherwise unconstrained,
see eqs. (\ref{tau}) and (\ref{tensor-K}).
Consider an off-shell tensor multiplet described by a symmetric 
real superfield $H^{ij}(z)$, 
\be
\cD^{(i}_\a H^{jk)} = {\bar \cD}^{(i}_\ad H^{jk)}=0~, 
\qquad H^{ij} =H^{ji}~, \qquad 
\overline{H^{ij}} = H_{ij}~.
\label{tensor-anal}
\ee
Associated with $H^{ij}(z)$ is the $O(2)$ multiplet 
$H^{++}(z,u^+) = H^{ij} (z) u^+_iu^+_j$.
We will assume $H^{ij}$ to be nowhere vanishing, 
\be
H^{ij}H_{ij} \neq 0~,
\label{vev}
\ee
the condition required of  a superconformal compensator.
Then,  the SU(2) gauge freedom can be partially fixed as 
\be
H^{ij} = -\frac{{\rm i}}{2} \,(\s_1)^{ij}  \,G ~, 
\qquad {\bar G} =G>0~, \qquad 
(\s_1)^{ij}= \left(
\begin{array}{rr}
0 ~ &  1  \\
1     ~ & 0  
\end{array}
\right) ~,
\label{tensor-g-c}
\ee 
which leaves an unbroken U(1) gauge symmetry.
To be consistent with the constraint (\ref{tensor-anal}), 
the SU(2) connection should be 
\be
\F^i_\a{}^{jk} ={\rm i} \, \S^i_\a  (\s_1)^{jk} 
+ \ve^{i(j} E^{k)}_\a \ln G ~, 
\ee
with $ \S^i_\a$ a U(1) connection.
We will give an application of  the gauge condition (\ref{tensor-g-c}) 
in the next subsection.

Let us denote by $\cP^{(0,4)}(u^-)$ the differential operator in the square 
brackets in  (\ref{Sfin-0}). Then the component action can be rewritten as 
\bea
S&=& \int\rd^4 x \,e
\oint_C {\rm d}  \mu^{(-2,-4)}\, \cP^{(0,4)}(u^-) \cL^{++}(z,u^+) \big|~.
\eea
Without loss of generality, we can assume the north pole of ${\mathbb C}P^1$, 
i.e. $u^{+i} \propto (0,1)$, 
to be  outside of the integration contour, hence $u^+$
can be represented as 
\bea
u^{+i} =u^{+\1}(1,\z) =u^{+\1}\z^i ~,\qquad 
\z^i=(1,\z)~, \qquad \z_i= \ve_{ij} \,\z^j=(-\z,1)~,
\label{north-chart}
\eea
with $\z$ the local complex coordinate for   ${\mathbb C}P^1$.
Now, the projective invariance, eqs.  (\ref{delta-u-}) and (\ref{ode}), 
can be used to set 
\be
u^-_i \equiv \hat{u}^-_i = (1,0) ~, \qquad   
\quad ~\hat{u}^{-i}=\ve^{ij }\,\hat{u}^-_j=(0,-1)~.
\label{fix-u-}
\ee   
${}$Finally, representing the Lagrangian in the form
\be
\cL^{++}(z,u^+) =  {\rm i}\, u^{+\1} u^{+\2}\,
\cL(z,\z) =  {\rm i} \big( u^{+\1} \big)^2 \z\, \cL(z,\z)~, 
\label{L++toL}
\ee
the action turns into 
\bea
S&=& -\int\rd^4 x \,e \,  \cP 
\oint_C \frac{ {\rm d} \z}{2\p {\rm i}} \, \z\, \cL(z,\z)\big|~, \qquad \quad 
\cP := \cP^{(0,4)}(\hat{u}^-) ~.
\label{Sfin-zeta}
\eea
The important point is that the operator $\cP$  is $\z$-independent, and therefore
its presence is not relevant when  evaluating the contour integral.
If the original Lagrangian, $\cL^{++}$, depends on matter superfields only, 
the contour integral in (\ref{Sfin-zeta}) corresponds to that arising in a rigid 
superconformal theory  \cite{K-hyper2}.

\subsection{Application I: Gauge invariance of the vector-tensor coupling}

Let $S(\cL^{++})$ denote the action (\ref{InvarAc}).
Consider  $\cL^{++}_{\rm v-t} = H^{++} V$, where $H^{++}(z,u^+)$ is
a tensor multiplet (or a real $O(2)$ multiplet), 
and $V(z,u^+) $ a real weight-zero tropical multiplet 
(see \cite{KLRT-M} for more detail). 
The latter describes a massless vector multiplet 
provided there is gauge invariance 
\be
\d V = \l + \tilde{\l} ~, 
\label{vector-g-i}
\ee
where $\l(z,u^+)$ is an arctic weight-zero multiplet, 
and $\tilde{\l} (z,u^+)$ its smile conjugate (see \cite{KLRT-M} for more detail). 
We can now prove that the functional $S (H^{++} V) $ 
is invariant under the gauge transformation (\ref{vector-g-i}).
It is sufficient to prove that 
\be
S(H^{++} \l) =0~, 
\ee
for an arbitrary arctic weight-zero superfield $\l(z,u^+)$.
The latter follows from the fact that the action (\ref{Sfin-0}) 
with $\cL^{++} = H^{++} \l$ has no pole in the complex $\z$-plane.

\subsection{Application II: The c-map}
In this subsection we would  like to give a curved superspace 
description for  the c-map \cite{cmap1,cmap2}.
The problem of developing a superspace description for the c-map 
 has  already been discussed in  \cite{RVV} (see also \cite{BS})  and \cite{NPV}. 
However, since no projective superspace formulation  
for 4D  $\cN=2$ matter-coupled supergravity was available at that time, 
the only possible approach to address the problem  was (i) to use the existence 
of  a one-to-one correspondence between $4n$-dimensional quaternionic K\"ahler spaces
and $4(n+1)$-dimensional hyperk\"ahler manifolds possessing a homothetic 
Killing vector, and the fact that such hyperk\"ahler spaces 
(or ``hyperk\"ahler cones'' \cite{deWRV})  are the target spaces for rigid $\cN=2$
 superconformal sigma models; 
and (ii) to construct an appropriate hyperk\"ahler cone
associated with a rigid superconformal model of $\cN=2$ tensor multiplets.
Now, we are in a position to overcome all the limitations of the earlier
works.

In accordance with \cite{RVV},  a tensor multiplet model  
corresponding to the c-map is described by the Lagrangian 
\be
\cL^{++} = \frac{1}{2{\rm i} \,H^{++}_0} \Big( F( H^{++}_{ I}) 
- \bar{F} ( H^{++}_{ I}) \Big)~, \qquad I=1, \dots , N+1~. 
\label{cmap1}
\ee
Here $ H^{++}_{ I}$ and $H^{++}_0$ are tensor multiplets, 
with $H^{++}_0$ obeying the constraint (\ref{vev}), 
and $F(z^I)$  is a holomorphic homogeneous function of second degree, 
$F(c\,z^I)= c^2F(z^I)$. 
Thus we have to consider the following action:
\bea
S&=&  {\rm Im} \int\rd^4 x \,e \,  \cP 
\oint_C \frac{ {\rm d} \z}{2\p {\rm i}} \, 
\frac{F\big(H_I (\z) \big)}{H_0(\z)}\Big|~, 
\label{cmap2}
\eea
where the superfields $H_I(\z)$ and $H_0(\z)$ are defined as 
\bea
H^{++}_I(z,u^+) &=&  {\rm i} \big( u^{+\1} \big)^2  H_I(z,\z)~, 
\qquad 
H_I(\z) = \F_I + \z G_I -\z^2 \bar{\F}_I~, 
\eea 
and similarly for $H_0(\z)$.

Before we start studying the curved-superspace action (\ref{cmap2}), it is worth 
giving some comments about its flat superspace version.
Let $\cP_{\rm flat} $ and $\cL_{\rm flat} $
be  the flat-superspace counterparts of  the operator $\cP$ (\ref{Sfin-zeta}) and the Lagrangian 
$\cL$ (\ref{L++toL}). We obviously have 
\bea
\cP_{\rm flat} =\frac{1}{16} (D^\1)^2 ({\bar D}^\1)^2 
= \frac{1}{16} (D^\1)^2 ({\bar D}_\2)^2~, 
\eea
with $D^i_\a$ and ${\bar D}^\ad_i$ the flat spinor covariant derivatives. 
It is easy to see that the flat-superspace version of the analyticity conditions (\ref{ana-introduction})
implies $({\bar D}^\ad_\1 +\z {\bar D}^\ad_\2 )\cL_{\rm flat}(\z)=0$, and thus for the 
rigid supersymmetric action $S_{\rm flat}$ we get 
\bea
S_{\rm flat} &=&  {\rm Im} \int\rd^4 x \,
\oint_C \frac{ {\rm d} \z}{2\p {\rm i}} \,  \cP_{\rm flat} 
\frac{F\big(H_I (\z) \big)}{H_0(\z)}\Big|
=  {\rm Im} \int\rd^4 x \,
\frac{(D^\1)^2 ({\bar D}_\1)^2}{16}
\oint_C \frac{ {\rm d} \z}{2\p {\rm i} \z^2} \,
\frac{F\big(H_I (\z) \big)}{H_0(\z)}\Big| \non \\
&=& {\rm Im} \int\rd^4 x \, {\rm d}^2\q {\rm d}^2 {} {\bar \q} 
\oint_C \frac{ {\rm d} \z}{2\p {\rm i} \z^2} \,
\frac{F\big(H_I (\z) \big)}{H_0(\z)}\Big|_{\q_\2={\bar \q}^\2=0} 
\label{cmap-flat}
\eea
The action obtained defines an $\cN=2$ supersymmetric 
  theory formulated in $\cN=1$ superspace. 
It is the  $\cN=2$ superconformal model which was studied in \cite{RVV,NPV}.

In \cite{RVV},  the problem of evaluating the contour integral 
in (\ref{cmap-flat}) was reduced 
to that  solved several years earlier in \cite{GHK} (see also \cite{deWRV})
for the case of the rigid c-map.
Specifically, Ro\v{c}ek et al. \cite{RVV} imposed the SU(2) gauge condition 
(\ref{tensor-g-c}) or, equivalently,  $H_0(\z) =  \z G_0 $, which 
essentially corresponds the rigid c-map (more precisely, 
$G_0=1$ in the case of the rigid c-map, but the presence of $G_0$ 
is irrelevant for computing the contour integral).    
The subtlety with the analysis in  \cite{RVV} is that their tensor multiplet model 
is rigid superconformal, and hence the SU(2) parameters are constant.\footnote{Actually, 
in the case of rigid $\cN=2$ supersymmetry, if a tensor multiplet is constrained as in 
eq. (\ref{tensor-g-c}), then it  is simply constant,  $G={\rm const}$.}

In our case, however, the SU(2) transformations are local, 
and it is legitimate
to choose the gauge condition (\ref{tensor-g-c}). 
As a result, the action  turns into 
\bea
S&=&  {\rm Im} \int\rd^4 x \,e \,  \cP \,
\frac{F\big(\F_I  \big)}{G_0}\Big|~
\label{cmap3}
\eea
provided the contour $C$ in (\ref{cmap2}) is taken to be a circle around 
the origin in $\mathbb C$. Still, the consideration given is not quite satisfactory, 
because of a special  gauge chosen.

{}Fortunately, there is no need to impose any SU(2) gauge condition in order to do the contour
integral in (\ref{cmap2}). Following the rigid supersymmetric analysis of \cite{NPV}, 
we represent 
\bea 
H_0(\z) = -{\bar \F}_0 \Big( \z- \z_+\Big) \Big( \z- \z_- \Big) ~, 
\qquad 
\z_\pm = \frac{1}{2 {\bar \F}_0} \Big(G_0 \mp  \sqrt{G^2_0 +4 |\F_0|^2}   \Big)~
\eea
and choose the contour $C$ in (\ref{cmap2}) to be a small circle around
the root $\z_+$. This leads to
\bea
S&=&  {\rm Im} \int\rd^4 x \,e \,  \cP \,
\frac{F\big(H_I (\z_+) \big)}{ \sqrt{G^2_0 +4 |\F_0|^2} }\,\Big|~.
\label{cmap4}
\eea
Since 
\be
\z_+ = - \frac{2\F_0}{ \big( G_0  + \sqrt{G^2_0 +4 |\F_0|^2}  \big) }
~\stackrel{\F_0 \to 0}{\longrightarrow}~ 0~,
\ee
the covariant action (\ref{cmap4}) reduces to (\ref{cmap3}) in the limit $\F_0 \to 0$.
In the flat superspace limit, we reproduce the results of \cite{RVV,NPV}.

\section{Chiral representation for the action principle}
\setcounter{equation}{0}

In this section we derive a new representation for the action 
principle  (\ref{InvarAc}) as an integral over the chiral subspace.

The covariantly chiral projector $\bar{\D}$ was defined in section 3, eq. (\ref{chiral-pr}).
It turns out that $\bar{\D}$ can be given an alternative representation.
It is
\bea
\bar{\D} \oint (u^+ \rd u^{+}) \,U^{(-2)}&=&
\frac{1}{16}  \oint \frac{(u^+ \rd u^{+})}{(u^+u^-)^2}
\Big( ({\bar \cD}^-)^2 +4\bar{S}^{--}\Big) 
\Big( ({\bar \cD}^+)^2 +4\bar{S}^{++}\Big) U^{(-2)}~,~~~
\label{chiralproj2}
\eea
with $U^{(-2)}(z,u^+)$ an arbitrary isotwistor superfield of weight $-2$
(see \cite{KLRT-M} for the definition of isotwistor supermultiplets, as 
well as Appendix B below). 
As before,  the constant isotwistor $u^-_i$ 
is chosen to be  linearly independent from $u^+_i$, $ (u^+u^-)\neq0$, 
but otherwise is completely arbitrary.
It is proved in Appendix C that 
that 
the right-hand side of (\ref{chiralproj2}) 
(i) remains invariant under 
arbitrary projective transformations (\ref{projectiveGaugeVar}); 
and (ii) is covariantly chiral.

Let us  transform  the action functional  (\ref{InvarAc})  by making use of 
eqs. (\ref{chiralproj1}) and (\ref{chiralproj2}):
\bea
S(\cL^{++})&=&
\frac{1}{2\pi} 
\int \rd^4 x \,{\rm d}^4\q{\rm d}^4{\bar \q}\,E
\oint (u^+ \rd u^{+})\,
\frac{W{\bar W}\cL^{++}}{(\S^{++})^2} 
\non \\
&=& \frac{1}{2\pi} \int {\rm d}^4x \,{\rm d}^4 \q \, \cE \, \bar{\D}
 \oint (u^+ \rd u^{+}) \frac{W{\bar W}\cL^{++}}{(\S^{++})^2} 
\non \\
&=& \frac{1}{32\pi} \int {\rm d}^4x \,{\rm d}^4 \q \, \cE \, 
 \oint \frac{(u^+ \rd u^{+})} {(u^+u^-)^2}
 \Big( ({\bar \cD}^-)^2 +4\bar{S}^{--}\Big) 
\Big( ({\bar \cD}^+)^2 +4\bar{S}^{++}\Big) \frac{W{\bar W}\cL^{++}}{(\S^{++})^2}
\non \\
&=& \frac{1}{8\pi} \int {\rm d}^4x \,{\rm d}^4 \q \, \cE \, W
 \oint \frac{(u^+ \rd u^{+})} {(u^+u^-)^2}
 \Big( ({\bar \cD}^-)^2 +4 \bar{S}^{--}\Big) 
\frac{\cL^{++}}{\S^{++}}~,
\label{action-proj-chiral}
\eea
where we have used eq. (\ref{Sigma}), the chirality of the vector multiplet strength,
${\bar \cD}^\ad_i W=0$, and the fact that $\cL^{++}$, $\S^{++}$ and 
${\bar \S}^{++}$  obey the constrains (\ref{ana-introduction}).
This result can be interpreted as  a coupling of two vector 
multiplets described by the covariantly chiral field strengths $W$ and $\mathbb W$, 
respectively.
\bea
S(\cL^{++})&=&-\int {\rm d}^4x \,{\rm d}^4 \q \, \cE \, W \,{\mathbb W}~, \non \\
{\mathbb W}  &=& -\frac{1}{8\pi}  \oint \frac{(u^+ \rd u^{+})} {(u^+u^-)^2}
 \Big( ({\bar \cD}^-)^2 +4 \bar{S}^{--}\Big) {\mathbb V}~, 
 \qquad {\mathbb V}:=  \frac{\cL^{++}}{\S^{++}}~.
\eea
The composite superfield  ${\mathbb V}$ introduced
can be interpreted as a  tropical prepotential for the vector multiplet 
described by $\mathbb W$.

Let us choose the Lagrangian in (\ref{action-proj-chiral}) to be $\cL^{++} = H^{++} \l$, 
where $H^{++}(z,u^+) $ is a tensor multiplet, and $\l(z,u^+) $ an arctic multiplet.
Since both $H^{++}$ and $\l$ are independent of the vector multiplet described by 
the strengths $W$ and $\bar W$, the latter can be chosen such that $\S^{++}=H^{++}$. 
 Then 
 \bea
S(H^{++} \l)&=& \frac{1}{8\pi} \int {\rm d}^4x \,{\rm d}^4 \q \, \cE \, W
 \Big( ({\bar \cD}^-)^2 +4\bar{S}^{--}\Big) 
 \oint \frac{(u^+ \rd u^{+})} {(u^+u^-)^2}\,\l (z,u^+)~.
\eea
We can now represent $u^{+i} $ in the form (\ref{north-chart}) and 
fix the projective invariance by choosing $u^-_i$ as in (\ref{fix-u-}).
\bea
S(H^{++} \l)&=& -\frac{1}{8\pi} \int {\rm d}^4x \,{\rm d}^4 \q \, \cE \, W
 \Big( ({\bar \cD}^\1)^2 +4\bar{S}^{\1 \1}\Big) 
 \oint {\rm d}\z\,
\l (z,\z)=0~,
\eea
since the integrand of the contour integral possesses no pole in the $\z$-plane.
This completes our second proof of the fact 
that the vector-tensor coupling 
$\cL^{++}_{\rm v-t} = H^{++} V$, with  $H^{++}(z,u^+)$ is
a tensor multiplet
and $V(z,u^+) $ the tropical prepotential of a vector multiplet,
is invariant under  the gauge transformations (\ref{vector-g-i}).

In ref. \cite{K-2008}, it was postulated that any chiral integral of the form 
\bea
S_{\rm c}= \int \rd^4 x \,{\rm d}^4\q \, \cE \, \cL_{\rm c} &+& {\rm c.c.}~,
\qquad {\bar \cD}_\ad \cL_{\rm c} =0 ~, 
\eea
can be represented as follows:
\bea
S_{\rm c}&=&\frac{1}{2\pi} \oint (u^+ \rd u^{+})
\int \rd^4 x \,{\rm d}^4\q {\rm d}^4{\bar \q}\, E\,
\frac{{ W}{\bar { W}}  \cL^{++}_{\rm c} }{({ \S}^{++})^2 }~, \non \\
\cL^{++}_{\rm c} &=&
 -\frac{1}{4} { V} \,\Big\{ \Big( (\cD^{+})^2+4{S}^{++}\Big) \frac{\cL_{\rm c}}{ W}
+\Big( (\cDB^{+})^2+4\bar{S}^{++}\Big)
\frac{{\bar \cL}_{\rm c} }{\bar { W}} \Big\}~,~~~~~~
\eea
with $V(z,u^+)$  a  tropical prepotential for the vector multiplet 
characterized by the field strength $ W$.
This assertion can now be  immediately proved with the aid of (\ref{action-proj-chiral}).
\\
%%%%%%%%%%%%%%%%%%%%%%%%%%%%%%%%%%%%%%%%%%%%%%%%%%
%%%%%%%%%%%%%%%%%%%%%%%%%%%%%%%%%%%%%%%%%%%%%%%%%%

\noindent
{\bf Acknowledgements:}\\
We are grateful to Ian McArthur for reading the manuscript.
This work was supported  in part by the Australian Research Council.
At  a final stage of this project, G.T.-M. was  supported 
 by the endowment 
 of the John S.~Toll Professorship, the University of 
 Maryland Center for String \& Particle Theory, and
 National Science Foundation Grant PHY-0354401.

%%%%%%%%%%%%%%%%%%%%%%%%%%%%%%%%%%%%%%%%%%%%%%%%%%
%%%%%%%%%%%%%%%%%%%%%%%%%%%%%%%%%%%%%%%%%%%%%%%%%%
\appendix

\section{Superspace geometry of conformal supergravity} 
\setcounter{equation}{0}
\label{SCG}

Consider a curved 4D $\cN=2$ superspace  $\cM^{4|8}$ parametrized by 
local bosonic ($x$) and fermionic ($\q, \bar \q$) 
coordinates  $z^{{M}}=(x^{m},\q^{\mu}_i,{\bar \q}_{\dot{\mu}}^i)$,
where $m=0,1,\cdots,3$, $\mu=1,2$, $\dot{\mu}=1,2$ and  $i=\1,\2$.
The Grassmann variables $\q^{\mu}_i $ and $\teb_{\dot{\mu}}^i$
are related to each other by complex conjugation: 
$\overline{\q^{\mu}_i}=\teb^{\dot{\mu}i}$. 
The structure group is chosen to be ${\rm SO}(3,1)\times {\rm SU}(2)$ \cite{Grimm,KLRT-M},
and the covariant derivatives 
$\cD_{{A}} =(\cD_{{a}}, \cD_{{\a}}^i,\cDB^\ad_i)$
have the form 
\bea
\cD_{{A}}&=&E_{{A}}
~+~\O_{{A}}
~+~\F_{{A}}~.
\label{CovDev}
\eea
Here $E_{{A}}=E_{{A}}{}^{{M}}(z)\pa_{{M}}$ 
is the supervielbein, with $\pa_{{M}}=\pa/\pa z^{{M}}$,
\bea
\O_{{A}}&=&\hf\O_{{A}}{}^{bc}M_{bc}=\O_{{A}}{}^{\b\g}\,M_{\b\g}
+{\bar \O}_{{A}}{}^{\bd\gd}\,\bar{M}_{\bd\gd}
\label{Lorentzconnection}
\eea
is the Lorentz connection,
\bea
\F_{{A}}=\F_{{A}}{}^{kl}J_{kl}~,~~~J_{kl}=J_{lk}
\label{SU(2)connection}
\eea
is the SU(2)-connection.
The Lorentz generators with vector indices ($M_{ab}=-M_{ba}$) and spinor indices
($M_{\a\b}=M_{\b\a}$ and ${\bar M}_{\ad\bd}={\bar M}_{\bd\ad}$) are related to each other 
by the rule:
$$
M_{ab}=(\s_{ab})^{\a\b}M_{\a\b}-(\tilde{\s}_{ab})^{\ad\bd}\bar{M}_{\ad\bd}~,~~~
M_{\a\b}=\hf(\s^{ab})_{\a\b}M_{ab}~,~~~
\bar{M}_{\ad\bd}=-\hf(\tilde{\s}^{ab})_{\ad\bd}M_{ab}~.
$$ 
The generators of SO(3,1)$\times$SU(2)
act on the covariant derivatives as follows:\footnote{In what follows, 
the (anti)symmetrization of $n$ indices 
is defined to include a factor of $(n!)^{-1}$.}
\bea
&{[}J_{kl},\cD_{\a}^i{]}
=-\d^i_{(k}\cD_{\a l)}~,
\qquad
{[}J_{kl},\cDB^{\ad}_i{]}
=-\ve_{i(k}\cDB^\ad_{l)}~, \non \\
&{[}M_{\a\b},\cD_{\g}^i{]}
=\ve_{\g(\a}\cD^i_{\b)}~,\qquad
{[}\bar{M}_{\ad\bd},\cDB_{\gd}^i{]}=\ve_{\gd(\ad}\cDB^i_{\bd)}~,
~~~
{[}M_{ab},\cD_c{]}=2\eta_{c[a}\cD_{b]}~,
\label{generators}
\eea
while 
${[}M_{\a\b},\cDB_{\gd}^i{]}=
{[}\bar{M}_{\ad\bd},\cD_{\g}^i{]}={[}J_{kl},\cD_a{]}=0$.
Our notation and conventions correspond to \cite{BK,KLRT-M}; they 
almost coincide with 
those used in \cite{WB} except for the normalization of the 
Lorentz generators, including a sign in the definition of  
the sigma-matrices $\s_{ab}$ and $\tilde{\s}_{ab}$.

The supergravity gauge group is generated by local transformations
of the form 
\be
\d_K \cD_{{A}} = [K  , \cD_{{A}}]~,
\qquad K = K^{{C}}(z) \cD_{{C}} +\hf K^{ c  d}(z) M_{c  d}
+K^{kl}(z) J_{kl}  ~,
\label{tau}
\ee
with the gauge parameters
obeying natural reality conditions, but otherwise  arbitrary. 
Given a tensor superfield $U(z)$, with its indices suppressed, 
it transforms as follows:
\bea
\d_K U = K\, U~.
\label{tensor-K}
\eea

The  covariant derivatives obey (anti-)commutation relations of the form:
\bea
{[}\cD_{{A}},\cD_{{B}}\}&=&
T_{ {A}{B} }{}^{{C}}\cD_{{C}}
+\hf R_{{A} {B}}{}^{{c}{d}}M_{{c}{d}}
+R_{ {A} {B}}{}^{kl}J_{kl}
~,
\label{algebra}
\eea
where $T_{{A} {B} }{}^{{C}}$ is the torsion, 
and $R_{{A} {B}}{}^{kl}$ and 
$R_{ {A} {B}}{}^{{c}{d}}$ 
constitute the curvature.
${}$The torsion is subject to the 
following constraints \cite{Grimm}:
\bea
&T_{\a}^i{}_{\b}^j{}^{c}=
T_{\a}^i{}_{\b}^j{}^{\g}_k=T_{\a}^i{}_{\b}^j{}_{\dot{\g}}^k
=T_{\a}^i{}^{\dot{\b}}_j{}^{\g}_k
=T_{a}{}_{\b}^j{}^c
=T_{ab}{}^{c}=0~,
\non\\
&T_{\a}^i{}^\bd_j{}^c=-2\ri\d^i_j(\s^{c})_{\a}{}^{\bd}~,~~~
T_{a}{}_{\b}^j{}^\g_k=\d^j_k\, T_{a\b}{}^\g~.
\label{constr-1}
\eea
Here we have omitted some constraints which follow by complex conjugation.
The algebra of covariant derivatives is \cite{KLRT-M}
\begin{subequations} 
\bea
\{\cD_\a^i,\cD_\b^j\}&=&
4S^{ij}M_{\a\b}
+2\ve^{ij}\ve_{\a\b}Y^{\g\d}M_{\g\d}
+2\ve^{ij}\ve_{\a\b}\bar{W}^{\gd\dd}\bar{M}_{\gd\dd}
\non\\
&&
+2 \ve_{\a\b}\ve^{ij}S^{kl}J_{kl}
+4 Y_{\a\b}J^{ij}~, 
\label{acr1} \\
%%%%%%%%%%%%%%%%%%%%%%%%%%%%%%%%%%%%%%%%%%%%%
\{\cDB^\ad_i,\cDB^\bd_j\}&=&
-4\bar{S}_{ij}\bar{M}^{\ad\bd}
-2\ve_{ij}\ve^{\ad\bd}\bar{Y}^{\gd\dd}\bar{M}_{\gd\dd}
-2\ve_{ij}\ve^{\ad\bd}{W}^{\g\d}M_{\g\d}
\non\\
&&
-2\ve_{ij}\ve^{\ad\bd}\bar{S}^{kl}J_{kl}
-4\bar{Y}^{\ad\bd}J_{ij}~,
\label{acr2} \\
%%%%%%%%%%%%%%%%%%%%%%%%%%%%%%%%%%%%%%%%%%%%%
\{\cD_\a^i,\cDB^\bd_j\}&=&
-2\ri\d^i_j(\s^c)_\a{}^\bd\cD_c
+4\d^{i}_{j}G^{\d\bd}M_{\a\d}
+4\d^{i}_{j}G_{\a\gd}\bar{M}^{\gd\bd}
+8 G_\a{}^\bd J^{i}{}_{j}~,
\label{acr3}
\eea
\bea
{[}\cD_a,\cD_\b^j{]}&=&
\ri(\s_a)_{(\b}{}^{\bd}G_{\g)\bd}\cD^{\g j}
+{\frac{\ri}2}\Big(({\s}_a)_{\b\gd}S^{jk}
-\ve^{jk}({\s}_a)_\b{}^{\dd}\bar{W}_{\dd\gd}
-\ve^{jk}({\s}_a)^{\a}{}_\gd Y_{\a\b}\Big)\cDB^\gd_k
\non\\
&&
+{\frac{\ri}2}\Big((\s_a)_{\b}{}^{\dd} T_{cd}{}_\dd^j
+(\s_c)_{\b}{}^{\dd} T_{ad}{}_\dd^j
-(\s_d)_{\b}{}^{\dd} T_{ac}{}_\dd^j\Big)M^{{c}{d}}
\non\\
&&
+{\frac{\ri}2}\Big((\ts_a)^{\gd\g}\ve^{j(k}\cDB_\gd^{l)}Y_{\b\g}
-(\s_a)_{\b\gd}\ve^{j(k}\cDB_{\dd}^{l)}\bar{W}^{\gd\dd}
-{\frac12}(\s_a)_\b{}^{\gd}\cDB_{\gd}^{j}S^{kl}\Big)
J_{kl}~,
\label{acr4}
\\
%%%%%%%%%%%%%%%%%%%%%%%%%%%%%%%%%%%%%%%%%%%%%
{[}\cD_a,\cDB^\bd_j{]}&=&
-\ri(\s_a)_\a{}^{(\bd}G^{\a\gd)}
\cDB_{\gd j}
+{\frac{\ri}2}
\Big(({\ts}_a)^{\bd\g}\bar{S}_{jk}
-\ve_{jk}({\s}_a)_\a{}^{\bd}{W}^{\a\g}
-\ve_{jk}({\s}_a)^{\g}{}_\ad\bar{Y}^{\ad\bd}\Big)
\cD_\g^k
\non\\
&&
+{\frac{\ri}2}\Big((\s_a)_\d{}^{\bd}T_{cd}{}^{\d}_{ j}
+(\s_c)_\d{}^{\bd}T_{ad}{}^{\d}_{ j}
-(\s_d)_\d{}^{\bd}T_{ac}{}^{\d}_{ j}\Big)
M^{cd}
\non\\
&&
+{\frac{\ri}2}
\Big(-(\s_a)^{\g}{}_\gd\d_{j}^{(k}\cD_{\g}^{l)}\bar{Y}^{\bd\gd}
-(\s_a)_{\g}{}^{\bd}\d_{j}^{(k}\cD_{\d}^{ l)}{W}^{\g\d}
+{\frac12}(\s_a)_\a{}^\bd\cD^{\a}_{ j}\bar{S}^{kl}\Big)
J_{kl}~,~~~
\label{acr5}
\eea
\end{subequations}
where 
\begin{subequations}
\bea
T_{ab}{}^\g_k&=&-{\frac14}(\ts_{ab})^{\ad\bd}\cD^\g_k\bar{Y}_{\ad\bd}
+{\frac14}(\s_{ab})^{\a\b}\cD^\g_kW_{\a\b}
-{\frac16}(\s_{ab})^{\g\d}\cD^l_\d\bar{S}_{kl}~,\\
%%%%%%
T_{ab}{}_\gd^k&=&-{\frac14}(\s_{ab})^{\a\b}\cDB_\gd^{ k}Y_{\a\b}
+{\frac14}(\ts_{ab})^{\ad\bd}\cDB_{\gd}^{ k}\bar{W}_{\ad\bd}
-{\frac16}(\ts_{ab})_{\gd\dd}\cDB^{\dd}_{l}S^{kl}~.
\eea
\end{subequations}
Here the real four-vector $G_{\a \ad} $,
the complex symmetric  tensors $S^{ij}=S^{ji}$, $W_{\a\b}=W_{\b\a}$, 
$Y_{\a\b}= Y_{\b\a}$ and their complex conjugates 
$\bar{S}_{ij}:=\overline{S^{ij}}$, $\bar{W}_{\ad\bd}:=\overline{W_{\a\b}}$,
$\bar{Y}_{\ad\bd}:=\overline{Y_{\a\b}}$ obey additional constraints implied 
by the Bianchi identities.
They comprise the dimension 3/2 identities \cite{Grimm,KLRT-M}:
\bea
\cD_{\a}^{k}S_{kl}+\cD^{\g}_{l}Y_{\g\a}&=&0~,
~~~
\cD_{\a}^{(i}S^{jk)}=\cDB_{\ad}^{(i}S^{jk)}=0~,
~~~
\cD_{(\a}^{i}Y_{\b\g)}=0~, ~~~
\cD_\a^i\bar{W}_{\bd\gd}=0~,~~~~~~
\label{Bianchi-3/2-1}
\\
\cDB^\ad_{k}\bar{S}^{kl}+\cDB_\gd^{l}\bar{Y}^{\gd\ad}&=&0~, ~~~
\cDB^{\ad}_{(i}\bar{S}_{jk)}=\cD^{\a}_{(i}\bar{S}_{jk)}=0~,~~~
\cDB^{(\ad}_{i}\bar{Y}^{\bd\gd)}=0~,~~~
\cDB^\ad_iW^{\b\g}=0~,~~~~~~~
\label{Bianchi-3/2-2}
\\
\cD_\a^iG_{\b\gd}&=&
-{1\over 4}\cDB_{\gd}^iY_{\a\b}
+{1\over 12}\ve_{\a\b}\cDB_{\gd l}S^{il}
-{1\over 4}\ve_{\a\b}\cDB^{\dd i}\bar{W}_{\gd\dd}~,
\label{Bianchi-3/2-3}
\\
\cDB^{\ad}_{i}G^{\g\bd}&=&
{1\over 4}\cD^{\g}_{i}\bar{Y}^{\ad\bd}
-{1\over 12}\ve^{\ad\bd}\cD^{\g l}\bar{S}_{il}
+{1\over 4}\ve^{\ad\bd}\cD_{\d i}{W}^{\g\d}~.
\label{Bianchi-3/2-4}
\eea

It should be remarked that the constraints (\ref{constr-1}) are  invariant under 
super-Weyl transformations generated by a covariantly chiral superfield $\s$
\bea
\cDB^\ad_i\s=0~.
\eea
The reader is referred to \cite{KLRT-M,KTM-4D-confFlat} for the transformation laws 
of the covariant derivatives and torsion superfields under the super-Weyl transformations.

\section{Projective supermultiplets}
\setcounter{equation}{0}

A projective supermultiplet of weight $n$,
$Q^{(n)}(z,u^+)$, is a scalar superfield that 
lives on  $\cM^{4|8}$, 
is holomorphic with respect to 
the isotwistor variables $u^{+}_i $ on an open domain of 
${\mathbb C}^2 \setminus  \{0\}$.
The variable $u^{+}_i $ are constant and invariant under the structure group action.
The projective supermultiplet of weight $n$ is characterized by the following conditions:\\
(i) it obeys the covariant analyticity constraints (\ref{ana-introduction});\\
(ii) it is  a homogeneous function of $u^+$ 
of degree $n$, that is,  
\be
Q^{(n)}(z,c\,u^+)\,=\,c^n\,Q^{(n)}(z,u^+)~, \qquad c\in \mathbb{C}^*~;
\label{weight}
\ee
(iii)  gauge transformations (\ref{tau}) act on $Q^{(n)}$ 
as follows:
\bea
\d_K Q^{(n)} 
&=& \Big( K^{{C}} \cD_{{C}} + K^{ij} J_{ij} \Big)Q^{(n)} ~,  
\non \\ 
K^{ij} J_{ij}  Q^{(n)}&=& -\frac{1}{(u^+u^-)} \Big(K^{++} D^{(-1,1)} 
-n \, K^{+-}\Big) Q^{(n)} ~, \qquad 
K^{\pm \pm } =K^{ij}\, u^{\pm}_i u^{\pm}_j ~,
\label{harmult1}   
\eea 
where
\bea
D^{(-1,1)}
=u^{-i}\frac{\partial}{\partial u^{+ i}} ~,
\qquad
D^{(1,-1)}=u^{+ i}\frac{\partial}{\partial u^{- i}} ~.
\label{5}
\eea
The transformation law (\ref{harmult1}) involves an additional isotwistor,  $u^-_i$, 
which is subject 
to the only condition $(u^+u^-) := u^{+i}u^-_i \neq 0$, and is otherwise completely arbitrary.
By construction, $Q^{(n)}$ is independent of $u^-$, 
i.e. $\pa  Q^{(n)} / \pa u^{-i} =0$,
and hence $D^{(1,-1)}Q^{(n)}=0$.
One can see that $\d_K Q^{(n)} $ 
is also independent of the isotwistor $u^-$, $\pa (\d_K Q^{(n)})/\pa u^{-i} =0$,
due to (\ref{weight}). 

More generally, a weight-$n$ isotwistor superfield $U^{(n)}(z,u^+)$ 
is defined to live on  $\cM^{4|8}$, be holomorphic with respect to 
the isotwistor variables $u^{+}_i $ on an open domain of 
${\mathbb C}^2 \setminus  \{0\}$,
 and satisfy the conditions (ii) and (iii).

The operators $\cD^+_{ \a} $ and   
${\bar \cD}^+_{\dot  \a}$
 obey the anti-commutation relations:
\bea
\{  \cD^+_{ \a} , \cD^+_{ \b} \}
=4\, Y_{\a \b}\,J^{++}
+4 \, S^{++}M_{\a \b}~, \qquad 
\{\cD_\a^+,\cDB_\bd^+\}=
8 \,G_{\a \bd} \,J^{++}~,
\label{analyt}
\eea
where 
$J^{++}:=u^+_i u^+_j \,J^{ij}$ and 
$S^{++}:=u^+_i u^+_j \,S^{ij}$. 
It follows  from (\ref{harmult1})
\bea
J^{++} \,Q^{(n)}=0~, \qquad J^{++} \propto D^{(1,-1)}~,
\label{J++}
\eea
and hence the covariant analyticity constraints (\ref{ana-introduction}) are consistent.

We refer the reader to \cite{KLRT-M,KTM-4D-confFlat} 
for a more complete analysis of projective supermultiplets 
including their super-Weyl transformation laws 
and the definition of the ``smile'' (or analyticity-preserving) conjugation.

\section{On the chiral projector }
\setcounter{equation}{0}

In this appendix we prove that the right hand side of (\ref{chiralproj2}) 
(i) is invariant under arbitrary projective transformations 
(\ref{re-scale-u+}), (\ref{delta-u-}) and (\ref{ode});
and (ii) is covariantly chiral.

The expression (\ref{chiralproj2}) 
is manifestly invariant under 
arbitrary re-scalings of $u^+$, eq. (\ref{re-scale-u+}),
as well as under the $\a$-transformations  (\ref{delta-u-}). 
It remains to check invariance under infinitesimal $\b$-transformations 
(\ref{delta-u-}), with the parameter $\b(t)$ constrained as in  (\ref{ode}).
Applying the $\b$-transformation gives
\bea
&&\d 
\Big( ({\bar \cD}^-)^2 +4\bar{S}^{--}\Big) 
\Big( ({\bar \cD}^+)^2 +4\bar{S}^{++}\Big) U^{(-2)}
= 4 \b D^{(-1,1)}\bar{S}^{++}
\Big((\cDB^+)^2+4\bar{S}^{++}\Big)
U^{(-2)}
~.~~~~~~~~~
\label{D1}
\eea
Then, using (\ref{ode}) and the identity 
\bea
{\rd\over \rd t}V^{(+2)}
&=&2{(\dt{u}^+u^-)\over (u^+u^-)}V^{(+2)}
-{(\dt{u}^+u^+)\over (u^+u^-)}D^{(-1,1)}V^{(+2)}~,
\eea
which holds for any isotwistor superfield $V^{(+2)}$ of weight $+2$,  
such as the superfield
$\bar{S}^{++} \big((\cDB^+)^2+4\bar{S}^{++}\big)U^{(-2)}$
appearing on the right of (\ref{D1}),
it follows that 
\bea
{\b}{(\dt{u}^+ u^{+})\over\upm^2}
D^{(-1,1)}V^{(+2)}&=&
-{\rd\over\rd t}\Big({\b\over \upm}V^{(+2)}\Big)~.
\eea
This indeed demonstrates that  the right hand side of (\ref{chiralproj2}) is projective invariant.

Now let us prove that the right hand side of (\ref{chiralproj2}) is covariantly chiral.
{}First of all, consider a
weight-zero isotwistor superfield $P (z,u^+)$ such that
\bea
\cDB^+_\ad P=0~. 
\eea
An example of such a superfield is
$\big( ({\bar \cD}^+)^2 +4\bar{S}^{++}\big) U^{(-2)}$.
Using the identities
\bea
J^{--}P&=&-\upm D^{(-1,1)}P~,
\\
\cDB_{\ad}^{-}\big((\cDB^{-})^2+4\bar{S}^{--}\big)&=&
-4\bar{S}^{--}\cDB^{\bd -}\bar{M}_{\ad\bd}
-4\bar{Y}_{\ad\bd}\cDB^{\bd -}J^{--}
-{4\over 3}(\cDB_{\ad q}\bar{S}^{q-})J^{--}~,~~~~~~~~~
\\
{[}\cDB_\ad^{+},({\bar \cD}^-)^2{]} P&=&
\Big{(}
4({\bar \cD}_\ad^{-}\bar{S}^{+-})
-4\upm\bar{Y}_{\ad\bd}\cDB^{\bd-} 
\non\\
&&
-4D^{--}\bar{S}^{++}{\bar \cD}_\ad^{-}
-2D^{--}({\bar \cD}_\ad^{-}\bar{S}^{++})
\Big{)}P
~, 
\eea
one can show that 
\bea
&&
\cDB_{\ad k}\Big( ({\bar \cD}^-)^2 +4\bar{S}^{--}\Big) 
P=
{1\over\upm}
\Big{[}
u^+_k\cDB_\ad^{-} -u^-_k\cDB_\ad^{+} \Big{]}
\Big( ({\bar \cD}^-)^2 +4\bar{S}^{--}\Big)  P
\non\\
&&=  {D^{(-1,1)}\over\upm} 
\Big(2u^-_k({\bar \cD}_\ad^{-}\bar{S}^{++})
+4u^-_k\bar{S}^{++}{\bar \cD}_\ad^{-}
+4u^+_k\upm\bar{Y}_{\ad\bd}\cDB^{\bd -}
-2u^+_k(\cDB_{\ad}^+\bar{S}^{--})
\Big)
P
~.~~~~~~~~~
\label{D.8}
\eea
Secondly, we notice that for any superfield $V_k^{(+2)}(z,u^+)$, 
which is homogeneous in $u^+_i$
of degree $+2$, the following identity  holds 
\bea
{(\dt{u}^+u^+)\over (u^+u^-)^3}D^{(-1,1)}V_k^{(+2)}
&=&
-{\rd\over \rd t}\Big({1\over \upm^2}V_k^{(+2)}\Big)~.
\label{D.9}
\eea
Using (\ref{D.8}) and (\ref{D.9}), one can then prove that 
\bea
&&
\cDB_{\ad k}
\oint \frac{(u^+ \rd u^{+})}{(u^+u^-)^2}
\Big( ({\bar \cD}^-)^2 +4\bar{S}^{--}\Big) 
\Big( ({\bar \cD}^+)^2 +4\bar{S}^{++}\Big) U^{(-2)} =0~.
\eea
As a result,  the right hand side of (\ref{chiralproj2}) is indeed covarianly chiral.

%%%%%%%%%%%%%%%%%%%%%%%%%%%%%%%%%%%%%%%%%%%%%%%%%%%
%%%%%%%%%%%%%%%%%%%%%%%%%%%%%%%%%%%%%%%%%%%%%%%%%%%

\end{document}